\documentclass[fleqn,usenatbib]{mnras}

\usepackage[T1]{fontenc}
\usepackage{ae,aecompl}

\usepackage{graphicx}	
\usepackage{subfig} 
\usepackage{amsmath}	
\usepackage{amssymb}
\newcommand{\halfspace}{\hspace{1pt}}

\newcommand{\Lya}{Ly$\alpha$}

\newcommand\HI{{\hbox{H\halfspace$\rm \scriptstyle I$}}}
\newcommand\HII{{\hbox{H\halfspace$\rm \scriptstyle II$}}}
\newcommand\HeI{{\hbox{He\halfspace$\rm \scriptstyle I$}}}
\newcommand\HeII{{\hbox{He\halfspace$\rm \scriptstyle II$}}}
\newcommand\HeIII{{\hbox{He\halfspace$\rm \scriptstyle III$}}}

\newcommand\HIs{{\rm H\halfspace\scriptscriptstyle I}}
\newcommand\HIIs{{\rm H\halfspace\scriptscriptstyle II}}

\newcommand\HeIs{{\rm He\halfspace\scriptscriptstyle I}}
\newcommand\HeIIs{{\rm He\halfspace\scriptscriptstyle II}}
\newcommand\HeIIIs{{\rm He\halfspace\scriptscriptstyle III}}

\newcommand\nhat{\mathbf{\hat n}}
\newcommand\lsim{~\lower.5ex\hbox{$\buildrel < \over \sim$}~}
\newcommand\gsim{~\lower.5ex\hbox{$\buildrel > \over \sim$}~}

\title[Numerical methods for IGM reionization]{A comparison of numerical methods for computing the reionization of intergalacitc hydrogen and helium by a central radiating source}
       
\author[]{Ka-Hou Leong$^{1}$\thanks{E-mail:\ KH.Leong@ed.ac.uk (KHL)},
        Avery Meiksin$^{1}$, Althea Lai$^{1}$, K.~H. To$^{2}$\\
        $^{1}$SUPA\thanks{Scottish Universities Physics Alliance},
	$^{1}$Institute for Astronomy, University of Edinburgh,
        Blackford Hill, Edinburgh\ EH9\ 3HJ, UK\\
        $^{2}$Department of Physics, University of Tokyo, 7 Chome-3-1 Hongo, Bunkyo City, Tokyo 113-8654, Japan}

\shortcites{}

\begin{document}

\date{Accepted . Received ; in original form }
\pagerange{\pageref{firstpage}--\pageref{lastpage}} \pubyear{2022}
\maketitle
\label{firstpage}

\begin{abstract}
We compare numerical methods for solving the radiative transfer equation in the context of the photoionization of intergalactic gaseous hydrogen and helium by a central radiating source. Direct integration of the radiative transfer equation and solutions using photon packets are examined, both for solutions to the time-dependent radiative transfer equation and in the infinite-speed-of-light approximation. The photon packet schemes are found to be  more generally computationally efficient than a direct integration scheme. Whilst all codes accurately describe the growth rate of hydrogen and helium ionization zones, it is shown that a fully time-dependent method is required to capture the gas temperature and ionization structure in the near zone of a source when an ionization front expands at a speed close to the speed of light. Applied to Quasi-Stellar Objects in the Epoch of Reionization (EoR), temperature differences as high as $5\times10^4$~K result in the near-zone for solutions of the time-dependent radiative transfer equation compared with solutions in the infinite-speed-of-light approximation. Smaller temperature differences are found following the nearly full photoionization of helium in gas in which the hydrogen was already ionized and the helium was singly ionized. Variations found in the temperature and ionization structure far from the source, where the gas is predominantly neutral, may affect some predictions for 21-cm EoR experiments.
\end{abstract}


\section{Introduction}
\label{sec:Intro}

Numerical simulations are now a staple method for providing detailed descriptions of the complex processes in virtually all areas of astrophysics. The range in physical processes treated has increased from gravity and gas dynamics to include magnetic fields, a host of atomic and molecular reaction networks and radiative transport.

Including radiative transport is in particular numerically challenging when the gas is not everywhere optically thick. In this case, the transport of radiation is not diffusive so that devising efficient methods to solve the full set of radiative transfer (RT) equations must be addressed, particularly when the mean free path of the radiation exceeds other important length scales in an application that must be spatially resolved. As radiative transport is a long-range effect in these applications, its introduction seriously hampers computations both because of the required increase in memory to represent the radiation field and, potentially, because of a severely curtailed time step.

Applications to galactic and cosmological structure formation for which radiative transport is essential include star formation and the impact of stars and Quasi-Stellar Objects (QSOs) on intergalactic gas, including the reionization of the intergalactic medium (IGM). Various approximation methods have been introduced into numerical simulations to solve the radiative transfer equations. Applications of the methods in numerical simulations are typically done in a post-processing stage, ie, on top of previously computed solutions to the gas dynamical equations. A comparison of many of these schemes to applications to static gas problems is presented in \citet{2006MNRAS.371.1057I}. These tests are confined to the photoionization of hydrogen. Whilst the results from the different schemes are generally in good agreement on the placement of the ionization front, with discrepancies limited to 5--10\%, a much broader range of differences were obtained for other quantities. The ionization fractions showed differences up to a factor of a few to several, and temperatures differed by up to a few tens of percent. Some of these differences do not necessarily reflect differences in the algorithms, but may be attributed in part to the different frequency ranges covered for the sources. Some of the differences, however, appear intrinsic to the schemes.

Fully coupled radiative hydrodynamical codes have also been formulated, but often for restricted applications, either through a suppression of one or more spatial dimensions or through approximations made to the radiative transfer equations. A comparison of some of these methods is presented in \citet{2009MNRAS.400.1283I}.

Whilst the reionization of intergalactic hydrogen has been addressed in a wide variety of simulations \citep[see][for a recent review]{2022arXiv220802260G}, the reionization of helium has received much less attention. Yet its solution is vital for understanding the temperature evolution and small scale structure of the IGM, which has been used to place constraints on various dark matter candidates \citep[eg][]{2016JCAP...08..012B, 2017PhLB..773..258G, 2017PhRvL.119c1302I, 2019MNRAS.484.4273L} and for placing limits on the neutrino mass \citep{2010JCAP...06..015V}. The temperature evolution of the IGM in turn places constraints on the nature, abundance and evolution of the sources that reionized the hydrogen and helium in the IGM \citep[eg][]{2002ApJ...567L.103T, 2007MNRAS.380.1369T, 2012MNRAS.419.2880B, 2016MNRAS.460.1885U, 2019ApJ...872...13W, 2020MNRAS.497..906K}. Unlike for hydrogen reionization, the two ionization states of helium result in a broadened singly-ionized (\HeII) zone. As intergalactic helium is detected through the \HeII\ \Lya\ absorption signature, it is important to obtain an accurate solution for this zone. Additional applications of helium reionization include the detailed structure of the proximity zones around QSOs for constraints on the reionization of both hydrogen and helium, which depend on the lifetime of the sources, the size of the ionized regions they produce and on the temperature of the ionized gas in the zone \citep[eg][]{2015ApJ...806..142Z, 2020MNRAS.493.1330D, 2021MNRAS.505.5084W}. The possibility of the detection of a 21-cm signal during the Epoch of Reionization (EoR) also requires understanding the heating of the still neutral gas by high energy photons, as only a small amount of heating is able to affect the absorption signal against the Cosmic Microwave Background (CMB), and even convert it into an emission signal \citep{2000ApJ...528..597T, 2019MNRAS.487.1101R, 2020ApJ...888..112M}.

In the context of IGM reionization simulations, two broadly different algorithmic approaches have been used to compute the photoionization driven by isolated radiation sources, one based on representing the radiation field as photon packets and the other based on a direct integration of the radiative transfer equation. Photon packet schemes have generally been used rather than direct integration for numerical simulations because of their greater numerical efficiency \citep{2004MNRAS.348L..43B}. Several three-dimensional numerical RT packages incorporating multi-frequency radiation have been used to compute the photoionization of both hydrogen and helium in the IGM, including \texttt{LICORICE} \citep{2010A&A...523A...4B}, \texttt{RADAMESH} \citep{2011MNRAS.411.1678C}, \texttt{TRAPHIC} \citep{2011MNRAS.412.1943P}, \texttt{C$^2$-RAY} \citep{2012MNRAS.421.2232F},  \texttt{CRASH} \citep{2013MNRAS.431..722G} and \texttt{RADHYDRO} \citep{2017ApJ...841...87L}.

With the exception of \texttt{TRAPHIC}, these codes provide only quasi-time-dependent solutions in the sense that they allow for evolution in the gas properties and the sources, but they solve only the static RT equation by taking the speed of light to be infinite. The infinite-speed-of-light approximation (ISLA) has two shortcomings:\ the ionization fronts may propagate to acausally large distances (greater than the distance light could travel), and the approximation is not valid when the ionization front is propagating near the speed of light. The propagation of superluminal ionization fronts is particularly a concern for the ionization zones produced by QSOs:\ because of their high luminosities, the ionization fronts may move at superluminal velocities over the lifetime of the sources, producing unphysically large ionization regions. This may be partly overcome by enforcing a light-horizon, eg, by removing photon packets when they exceed their causal horizon. Such a solution, however, does not conserve the radiative energy carried by the photons and may alter the post-ionization temperature, and therefore ionization fractions, of the gas. The computation of ionization fronts propagating at near the speed of light requires an algorithm that solves the time-dependent radiative transfer equation to obtain an accurate post-ionization temperature, as will be demonstrated in this paper. One goal of this paper is to provide a means for implementing an approximate practical photon packet RT scheme for cosmological simulations that handles near-luminal ionization front expansion without solving the time-dependent RT equation, which is prohibitively computationally expensive.

Photon packet schemes have been applied to the reionization of intergalactic hydrogen and helium both as a post-processing step \citep{2002MNRAS.332..601S, 2004MNRAS.348L..43B, 2009ApJ...694..842M, 2012MNRAS.423..558C, 2013MNRAS.435.3169C, 2017MNRAS.468.3718K, 2018MNRAS.476.1174E, 2020MNRAS.498.6083E} and in fully coupled radiative hydrodynamics (RHD) schemes \citep{2012MNRAS.423....7M, 2017ApJ...841...87L}. Moments based schemes for solving the radiative transfer equations to study the reionization of hydrogen and helium in the IGM have been employed both in a multi-step scheme \citep{2022arXiv220713098P} and in RHD simulations \citep{2019MNRAS.490.3177W, 2022MNRAS.511.4005K}. Codes that compute the reionization of hydrogen and helium by a QSO in a static (or comoving) gas and assuming spherical symmetry have been used to estimate the gas properties around a QSO using both direct integration \citep{1997ApJ...475..429M, 2000ApJ...528..597T, 2012MNRAS.421.2232F} and photon packet schemes \citep{2008ApJ...686..195Z, 2016MNRAS.457.3006D, 2016ApJ...824..133K, 2018MNRAS.479.4320G, 2021ApJ...917...38E, 2021ApJ...921...88M}.

Rather than comparing results from a broad range of numerical codes, the focus of this paper is on the requirements for achieving convergent results using different numerical RT methods, with a particular view to application to the reionization of helium in the IGM. We compare two photon packet algorithms and a direct integration algorithm applied to sources with spectra similar to stars and QSOs. Whilst previous convergence tests focused on the extent of ionization zones and their properties using solutions of the time-independent RT equation (ISLA methods), special attention is given here to differences arising between solutions to the time-dependent (for which the speed of light is finite) and time-independent RT equations. In addition to eliminating spurious fast-than-light growth of the ionization zones, we show there are significant differences in the post-ionization temperature in the near-zone of QSO sources when solving the time-dependent RT equation compared with ISLA solutions. With a view to applications to the 21-cm EoR signal, we also examine convergence in the far-zone, where the gas is still largely neutral. We focus on idealised test problems, as these provide the best means for isolating differences in the behaviour of the algorithms. To demonstrate the consequences of these differences in a cosmological setting, we also provide results for the ionization by a source in a cosmological simulation. In the next section, we describe the numerical methods. In Sec.~\ref{sec:tresults}, test problems are described and results presented. These are discussed in Sec.~\ref{sec:discussion}, where also results for a cosmological simulation are presented, and conclusions are summarised in Sec.~\ref{sec:conclusions}. Comparisons between the time-dependent solutions to the RT equation and published ISLA solutions for test problems are presented in Appendix~\ref{appendix:comparisons}. One of the photon packet schemes is a module of the gravity-hydrodynamics code \texttt{ENZO}. Convergence test results for \texttt{ENZO} are provided in Appendix~\ref{appendix:convergence}. Alterations to the code required to carry out the tests are described in Appendex~\ref{appendix:enzo_revision}. 

\section{Numerical Solutions of the Reionization Equations}
\label{sec:NSrieqs}

\subsection{The Reionization Equations}
\label{subsec:Rieqs}

The ionization equations for hydrogen and helium are

\begin{eqnarray}
\frac{dx_\HIs}{dt} &=& -x_{\rm HI}\left[\Gamma_\HIs +
                           n_e\gamma_\HIs(T)\right]
+x_\HIIs n_e\alpha_\HIIs(T), \nonumber\\
\frac{dx_\HIIs}{dt} &=& -\frac{dx_\HIs}{dt}, \nonumber\\
\frac{dx_\HeIs}{dt}  &=& -x_\HeIs \left[\Gamma_\HeIs
                         +n_e\gamma_\HeIs(T)\right] + x_\HeIIs n_e\alpha_\HeIIs, \nonumber \\
\frac{dx_\HeIIs}{dt}&=& -\frac{dx_\HeIs}{dt} - \frac{dx_\HeIIIs}{dt}, \nonumber \\
 \frac{dx_\HeIIIs}{dt}&=&  x_\HeIIs \left[\Gamma_\HeIIs +
                          n_e\gamma_\HeIIs(T)\right] - x_\HeIIIs
                          n_e\alpha_\HeIIIs,
\label{eq:phionize}
\end{eqnarray}
where $\alpha_\HIIs$, $\alpha_\HeIIs$ and $\alpha_\HeIIIs$ are the
total radiative recombination rates to all levels of \HI, \HeI\ and \HeII, respectively,  $\Gamma_\HIs$,
$\Gamma_\HeIs$ and $\Gamma_\HeIIs$ are the corresponding photoionization rates, and $\gamma_\HIs(T)$,
$\gamma_\HeIs(T)$ and $\gamma_\HeIIs(T)$ are the corresponding collisional
ionization coefficients. For hard spectra, secondary ionizations produced by ejected electrons may help to partially ionize the gas. We do not include this effect to focus comparisons between the codes on the numerical solutions of the radiative transfer equations; including secondary electron ionizations could in principle complicate the interpretation of any differences. Their correct implementation is moreover hampered by the long path lengths of the secondary electrons compared with the length scales of interest. A comparison between different treatments is provided by \citet{2016MNRAS.457.3006D}. To illustrate the magnitude of the effects of secondary electron ionizations, we present some results for test problems without and with secondary electron ionizations in Appendix~\ref{appendix:comparisons}. Generally the effects are only moderate, but warrant inclusion for more precise solutions; the effects of secondary electron ionizations are particularly large in the partially ionized regions outside the main ionized zones. The respective \HI, \HII, \HeI, \HeII\ and \HeIII\ fractions are denoted by $x_{\rm HI}$, $x_{\rm HII}$, $x_{\rm HeI}$, $x_{\rm HeII}$ and $x_{\rm HeIII}$. The total electron density is $n_e = n_{\rm HII} + n_{\rm HeII}
+ 2n_{\rm HeIII}$. The time-derivatives are lagrangian, so that Eqs.~(\ref{eq:phionize}) are valid in the presence of velocity flows.

The photoionization rate per atom (or ion) of a species $i$ (\HI, \HeI\ or \HeII) is
\begin{equation}
\Gamma_i = c\int_{\nu_{{\rm T}, i}}^\infty\,d\nu
\frac{u_\nu}{h\nu}a_{i, \nu},
\label{eq:Gion}
\end{equation}
where $a_{i,\nu}$ is the photoelectric cross section of species $i$,
$\nu_{{\rm T}, i}$ is the threshold frequency required to ionize
species $i$, $u_\nu$ is the specific energy density of the
ambient radiation field, and $h$ is Planck's constant. The specific energy density is related to the specific intensity of the radiation field $I_\nu({\bf r},t,{\hat{\bf
n}})$ by $u_\nu=4\pi J_\nu/c$, where $J_\nu({\bf
r},t)=(1/4\pi)\oint\,d{\bf\Omega} I_\nu(\mathbf{r},t,\nhat)$ is
the angle-averaged specific intensity at position $\mathbf{r}$ at time
$t$ in the direction $\nhat$.

The equation of radiative transfer for $I_\nu({\bf r},t,\nhat)$ in a static medium with absorption coefficient $\alpha_\nu({\bf r},t,\nhat)$ and emission
coefficient $j_\nu({\bf r},t,\nhat)$ is
\begin{eqnarray}
\frac{1}{c}\frac{\partial I_\nu({\bf r},t,\nhat)}{\partial t}+\nhat\cdot
{\bf \nabla}I_\nu({\bf r},t,\nhat) = \nonumber\\
-\alpha_\nu({\bf r},t,\nhat)
I_\nu({\bf r},t,\nhat) + j_\nu({\bf r},t,\nhat).
\label{eq:tdRT}
\end{eqnarray}
In the context of reionization, the emission coefficient
represents a source like a star, galaxy or QSO, although in principle
it may also account for photoionizing radiation following radiative
recombination within the gas. Since only continuum radiation is considered, as distinct from resonance lines, the static medium approximation is adequate on scales small compared with the cosmic horizon.

The value of $I_\nu$ at any given time $t$ and position $s$ along the
direction $\nhat$ will be given by any incoming intensity $I_\nu^{\rm
inc}$ at position $s_0$ at time $t_{\rm ret}=t-(s-s_0)/c$, absorbed
by intervening material at positions $s^\prime$ at the retarded times
$t^\prime_{\rm ret}=t-(s-s^\prime)/c$, along with contributions from
sources at positions $s^{\prime\prime}$ that emitted at the retarded
times $t^{\prime\prime}_{\rm ret}=t-(s-s^{\prime\prime})/c$, followed
by absorption. Accordingly, the formal solution to Eq.~(\ref{eq:tdRT}) is
\begin{eqnarray}
I_\nu(s,t) = \left(I_\nu^{\rm inc}\right)_{s_0,t_{\rm ret}}
\exp\left[-\int_{s_0}^s\,ds^\prime\, 
(\alpha_\nu)_{s^\prime,t_{\rm ret}^\prime}
\right]\nonumber\\
+\int_{s_0}^s\,ds^{\prime\prime}\, 
\left(j_\nu\right)_{s^{\prime\prime},t_{\rm ret}^{\prime\prime}}
\exp\left[-\int_{s^{\prime\prime}}^s\,ds^\prime\, 
(\alpha_\nu)_{s^\prime,t_{\rm ret}^\prime}
\right]. 
\label{eq:tdRT-sol}
\end{eqnarray}
We shall refer to such solutions as solutions to the \emph{time-dependent} RT equation. By contrast, most of the literature makes the infinite-speed-of-light approximation, which corresponds to solving the \emph{time-independent} (or static) RT equation, for which the term involving the time-derivative of $I_\nu$ in Eq.~(\ref{eq:tdRT}) is absent, and only the instantaneous properties of the gas appear in Eq.~(\ref{eq:tdRT-sol}) rather than the time-retarded properties. The solutions may none the less be quasi-time-dependent when using the ISLA if the properties of the gas and the source vary with time, but only on timescales long compared with the light propagation time from the source.

The photoionization will also heat the gas, while radiative
recombination and collisional effects will cool the gas. We follow the
approach outlined in \citet{2009RvMP...81.1405M}. Specifically, we solve the energy equation in the form

\begin{equation}
    \frac{dS_E}{dt} = (\gamma-1)\rho^{-\gamma}\left(G-L\right),
    \label{eq:dSEdt}
\end{equation}
where $S_E = p/\rho^\gamma$ for gas pressure $p$, mass density $\rho$ and ratio of specific heats $\gamma$. Here, $G$ is the heating rate per unit volume of the gas and $L$ is the radiative energy loss rate per unit volume. The photoionization heating rate per volume for a species $i$ of number density $n_i$ is

\begin{equation}
G_i = n_ic\int_{\nu_{\rm T}, i}^\infty\,d\nu
\frac{u_\nu}{h\nu}a_{i, \nu}h\left(\nu - \nu_{{\rm T}, i}\right).
\label{eq:GHion}
\end{equation}
The total heating rate from all species is $G=G_\HI + G_\HeI + G_\HeII$. The radiative energy loss term $L$ includes energy losses from radiative recombination, collisional excitation and inverse Compton cooling off the Cosmic Microwave Background, using the rates referenced in \citet{2009RvMP...81.1405M}. For applications in Appendix~\ref{appendix:comparisons}, energy losses from secondary electron ionizations and adiabatic cooling from cosmic expansion (allowing for an evolving mass density) are included for some of the test problems. The gas temperature is computed from
\begin{equation}
    T = \frac{\bar m}{k}S_E\rho^{\gamma-1},
    \label{eq:T}
\end{equation}
where $\bar m$ is the mean mass per particle and $k$ is Boltzmann's constant. The ionization scenarios treated are idealized in that they do not allow for additional energy losses from dust or metals, as may occur in the ionized regions of high redshift QSO sources. Modelling such effects is well beyond the intent of this paper and would only complicate the interpretation of differences in the results arising from the different photoionization algorithms considered.


\subsection{Methods of Numerical Solution}
\label{subsec:meths}

We confine the discussion to algorithms that solve the radiative transfer equation along individual rays, as distinct from moment methods. Full 3D RT is achieved through the means of constructing the rays, a topic we shall not discuss, simply adopting the existing framework for the 3D code used (\texttt{ENZO}). We consider two types of numerical methods for solving the 1D RT equation, one based on a direct
integration of the time-independent RT equation (an ISLA method) and the other for which the radiation is represented by photon packets. Two versions of the latter are tested, one corresponding to integrating the time-independent RT equation (an ISLA method) and the other corresponding to integrating the time-dependent RT equation (retaining the differential time operator). Each method is outlined in turn.

\subsubsection{Instantaneous direct integration}
\label{subsubsec:idi}

At each time step, the 1D radiative transfer equation is integrated along
the line of sight. Retaining the past absorption coefficient for all
positions along the line of sight at previous times is normally
prohibitively expensive in a simulation, so instead new rays are cast
for each time step and the absorption coefficient for that time step is
used. The solution is taken as
\begin{eqnarray}
I_\nu(s,t) = \left(I_\nu^{\rm inc}\right)_{s_0,t}
\exp\left[-\int_{s_0}^s\,ds^\prime\, 
(\alpha_\nu)_{s^\prime,t}
\right]\nonumber\\
+\int_{s_0}^s\,ds^{\prime\prime}\, 
\left(j_\nu\right)_{s^{\prime\prime},t}
\exp\left[-\int_{s^{\prime\prime}}^s\,ds^\prime\, 
(\alpha_\nu)_{s^\prime,t}
\right]. 
\label{eq:tdRT-appsol}
\end{eqnarray}
This in effect treats the speed of light as infinite, although a
cut-off in the distance the radiation reaches is imposed to ensure
the extent of the region affected by a source preserves
causality. This instantaneous solution may be a good approximation
when the ionization front moves much faster than the gas flows. In the
implementation used here, a spatial grid is used along the line of
sight, with the width of each zone chosen to ensure the optical depth
of still neutral hydrogen or singly ionized helium is approximately 1. This ensures an accurate integration of the radiative transfer equation. The time step is chosen
to be a fraction of the shortest ionization or cooling time,
sufficient to provide convergence at the few percent level. The energy equation is solved with an explicit second order time integrator, and an implicit scheme is used to solve the time-dependent ionization equations \citep{1994ApJ...431..109M}.


\subsubsection{Rays and photon packets}
\label{subsubsec:raypp}

An alternative approach traces packets of photons of distinct energy
groups along rays. The algorithm consists of two parts:\ casting rays
through a simulation volume from each source, and propagating photon
packets along the rays. For more than a single spatial dimension, rays bunch together near a source and the separations between the rays increase with distance.
Adequate cell coverage of the regions affected by each source is assured by splitting rays as required, as in \citet{2002MNRAS.330L..53A}. Photon packets are then propagated along the rays on each time step, until the packets are either completely
absorbed or escape the simulation volume.

The gas component in the simulations is computed on a spatial grid. The
fraction of the photon packets absorbed on crossing a grid zone
depends on the optical depth through the zone. An advantage of this
method over the direct integration scheme above is that the optical
depth may be chosen to be above 1 while maintaining good accuracy; up
to 10 is typical, but even larger values are able accurately to recover the expansion of an ionization front \citep{1999ApJ...523...66A}. This considerably reduces the computational time. A consequence, however, is an ambiguity in how to share the photons when more than a single species may be ionized in a zone, as they will
compete for the same photons. How the gas is ionized in a
zone, were it completely resolved, can change the probability for
photons of different energies to be absorbed. In the simplest implementation, such as in \texttt{ENZO}, the fraction of photons of energy $i$ absorbed by
species $j$ is $1-\exp[-\tau_j(\nu_i)]$, but in looping over $j$,
different fractions may result depending on how the species are
ordered in the loop. We adopt instead the more balanced probabilistic
approach of \citet{2004MNRAS.348L..43B}, and have introduced this into the version of \texttt{ENZO} we use. The absorption probabilities
of a photon in a packet of frequency $\nu$ by \HI, \HeI\ and \HeII\ are respectively

\begin{equation}
 P_{\rm abs}^{\rm HI} = p_{\rm HI}q_{\rm HeI}q_{\rm HeII}\large[1-\exp(-\tau_{\nu}^{\rm total})\large]/D,
	\label{eq:abs_HI}
\end{equation}
\begin{equation}
 P_{\rm abs}^{\rm HeI} = q_{\rm HI}p_{\rm HeI}q_{\rm HeII}\large[1-\exp(-\tau_{\nu}^{\rm total})\large]/D,
	\label{eq:abs_HeI}
\end{equation}
\begin{equation}
 P_{\rm abs}^{\rm HeII} = q_{\rm HI}q_{\rm HeI}p_{\rm HeII}\large[1-\exp(-\tau_{\nu}^{\rm total})\large]/D.
	\label{eq:abs_HeII}
\end{equation}
Here $p_{i} = 1 - \exp(-\tau_{\nu}^{\rm i})$ and $q_{i} =
\exp(-\tau_{\nu}^{\rm i})$ are the auxiliary absorption and
transmission probabilities for the species $i$, $D = p_{\rm HI}q_{\rm
  HeI}q_{\rm HeII} + q_{\rm HI}p_{\rm HeI}q_{\rm HeII} + q_{\rm
  HI}q_{\rm HeI}p_{\rm HeII}$ is the normalisation factor, and
$\tau_{\nu}^{\rm total} = \tau_{\nu}^{\rm HI} + \tau_{\nu}^{\rm HeI}
+ \tau_{\nu}^{\rm HeII}$ is the total optical depth, where
$\tau_{\nu}^{i}$ is the optical depth of species $i$.\footnote{In practice, for the test problems presented here, the different ionization zones are sufficiently distinct that the results are nearly the same using a simpler formulation treating the absorption by each species independently, with results agreeing typically to better than a percent. Differences up to 20 percent, however, may arise for low ionization fractions in some regions. We retain the formulation described here for generality.}

We test the implementation of the photon packet scheme used in the numerical-hydrodynamics code \texttt{ENZO} v2.6 \citep{2011MNRAS.414.3458W,  2014ApJS..211...19B}, modified as described in Appendix~\ref{appendix:enzo_revision}, including the photon absorption probabilities above, with photoionization cross-sections from \citet{1997NewA....2..209A}\footnote{The \HeI\  photoionization cross-section is updated to that of \citet{1996ApJ...465..487V}.} , and the chemistry and cooling solver \textbf{GRACKLE} \footnote{https://grackle.readthedocs.io/}\citep{2017MNRAS.466.2217S}. Because the code runs in 3D, the memory cost of retaining photon packets from previous time steps rapidly becomes prohibitive in reionization problems. For this reason, instantaneous radiative transfer is assumed in the code (ISLA). New packets are generated on each time step. This corresponds to neglecting the time derivative in Eq.~(\ref{eq:tdRT}), or equivalently adopting an infinite speed of light. To preserve causality, we also permanently delete any surviving photon packets that travel outside the light cone of the source (see Appendix~\ref{appendix:horizon}.) This necessarily results in a loss of radiant energy and an artificially reduced photoionization heating rate of the gas.

For a more physical contrast compared with the other two methods, we also present results using a 1D spherically symmetric code, \texttt{PhRay}, that propagates the photon packets at a finite velocity. The photon packets are kept between time steps. They then have a memory of the gas they passed through on all previous time steps. This coresponds to retaining the time derivative in Eq.~(\ref{eq:tdRT}). Adopting the speed of light for the packet velocity requires very short time steps, as the code is written to ensure photons cannot move more than a single grid zone in one time step. (A
more efficient code could be written to allow movement through multiple grid zones, but this would entail considerable additional computational overheads.) The grid zones are adjusted to a preset maximum optical depth per zone. In outline, the basic steps of the code are:

\begin{enumerate}

\item Choose the length of the ray.
\item Compute the minimum cell width to ensure the \HI, \HeI\ and \HeII\ optical depths do not exceed preset maximum values for a cell. Set the time step to the time it takes a photon to cross a single cell.
\item On the first time step, add photon packages to the first cell nearest the source according to the source luminosity and time step.
\item On subsequent time steps, move photon packages in each cell to the next cell away from the source. Solve the ionization equations Eqs.~(\ref{eq:phionize}) and energy equation Eq.(\ref{eq:dSEdt}) in each cell. Remove photon packages according to the optical depth in the new cell, using Eqs.~(\ref{eq:abs_HI}) - (\ref{eq:abs_HeII}). Add new photons to the first cell nearest the source.
\item Repeat step (iv) until the final integration time.
    
\end{enumerate}

As long as the advance of the fastest ionization front is highly
sub-luminal, nearly identical results are produced allowing for a packet
velocity smaller than the speed of light, provided the packets
still move quickly compared with the ionization fronts. This has the advantage of allowing the code to take longer time steps. In our tests, we use the physical speed of light.

For both photon packet codes, the energy and ionization equations are solved using a first order Eulerian time integrator.

\subsection{Frequency integration}
\label{subsec:freqint}

For the photon packet schemes, the integration of the radiation field over the photoionization cross-section to compute the ionization and heating rates is accomplished using Gaussian Legendre quadrature. This is more efficient than a uniform grid in frequencies, and yields extremely accurate integrations when the integrand is well-approximated by a polynomial. We experimented with several implementations, and use one we find ensures accuracy in the frequency integrations typically to within a percent for the applications we present. Specifically, the frequency range is divided into intervals between the photoelectric edges for \HI\ ($13.6\,\mathrm{eV}$), \HeI\ ($24.6\,\mathrm{eV}$) and \HeII\ ($54.4\,\mathrm{eV}$), and extending to an upper limiting value dependent on the RT method used and the application. For the spherically symmetric photon packet method (\texttt{PhRay}), 8 energy bins were used in each frequency interval. The integration variable for the final integration to infinite frequency is changed to $1/\nu$, so that the integration ranges from $0$ to $1/\nu_{L, \HeII}$, where $\nu_{L, \HeII}$ is the ionization threshold for \HeII. For the \texttt{ENZO} simulations, instead an upper frequency is imposed, provided in the test problem descriptions below. The number of frequency bins between threshold energies is set at 5, with 10 between the \HeII\ photoelectric threshold and the maximum energy. The energy binnings used for the photon packet codes ensure accuracy in the frequency integrations to within a percent.

For the direct integration method, Gaussian quadrature offers some advantage over a mixed linear-logarithmic frequency grid in allowing a reduction by a factor of three in the number of frequencies used for an accurate solution within the ionized region, however the ionization front moves somewhat too quickly unless a comparable number of frequencies is used. We consequently use a mixed linear-logarithmic frequency grid for the direct integration scheme, with the number of frequencies typically 200, half placed uniformly between the \HI\ and \HeII\ photoelectric thresholds and the remainder placed logarithmically to a maximum energy of 2~keV. This choice ensures temperatures and placements of ionization fronts are converged to better than a percent at a given spatial grid resolution.

\section{Test results}
\label{sec:tresults}

\subsection{Test problems}
\label{subsec:tproblems}

The algorithms are tested on four problems, reionization by a black body spectrum with a temperature of $10^5$~K or $10^6$~K, and reionization under intergalactic medium conditions by a power-law spectrum before and after hydrogen is ionized. The $10^5$~K black body roughly represents a high mass Pop~II or Pop~III star \citep{1984ApJ...280..825B}, whilst the $10^6$~K black body (hotter than expected for stellar atmospheres), is presented as a contrasting source with a greater proportion of ionizing photons able to fully ionize helium. The power-law spectra represent QSOs, although they also approximate the photon emission rates for galaxies dominated by Pop~II or Pop~III stars \citep{2005MNRAS.356..596M}. The mass-fraction abundances of hydrogen and helium are $0.76$ and $0.24$, respectively. In all cases, the surrounding gas is static and its hydrodynamical response is not included, since the focus is on the ionization structure. The problems and results are discussed in greater detail below.

\subsection{Black body spectra}
\label{subsec:bbspec}

The luminosity function of the black body spectrum is modelled as 
\begin{equation}
   L_{BB,\nu}=L_{0}\times\frac{(h \nu)^{3}}{\exp{\frac{h \nu}{k T}}-1}\,\mathrm{eV s^{-1} Hz^{-1}}.
   \label{eq:bb}
\end{equation}
 The surrounding hydrogen number density is $0.76\times10^{-3}\mathrm{cm}^{-3}$. Hydrogen and helium are initially neutral in all simulations. The physical boxsize in all the \texttt{ENZO} 3D instantaneous simulations with $256^{3}$ cells is $6.6\rm kpc$.\footnote{The parameters are adopted from the {\it PhotonTest} test problem in \texttt{ENZO} v2.6.} The optical depths of all species per zone are below 1.
The total number of cells in the 1D simulations is adjusted in different situations, ensuring the optical depth of neutral hydrogen in each cell at the start of the computation is approximately equal to 1. This ensures the positions of the ionization fronts are converged to within a few percent.

We show that in the two black body radiation problems, all algorithms show consistent temperature patterns and the differences in the gaseous ionisation levels are negligible for practical applications.  

\subsubsection{$T_\mathrm{BB}=10^5$~K}
\label{subsubsec:PopII}
The initial gas temperature is $T=100$~K. The coefficient 
$L_{0} = 1.9\times10^{31}\,\mathrm{eV^{-2}}$
corresponds to a photon emission rate above the hydrogen ionization threshold of $\dot N_{\mathrm{H}, \gamma} = 5\times10^{48}\,\mathrm{s^{-1}}$. For the \texttt{ENZO} computations, the maximum energy bin used is $200\, \mathrm{eV}$ to ensure convergence on the temperature.

\begin{figure}
    \includegraphics[width=\columnwidth]{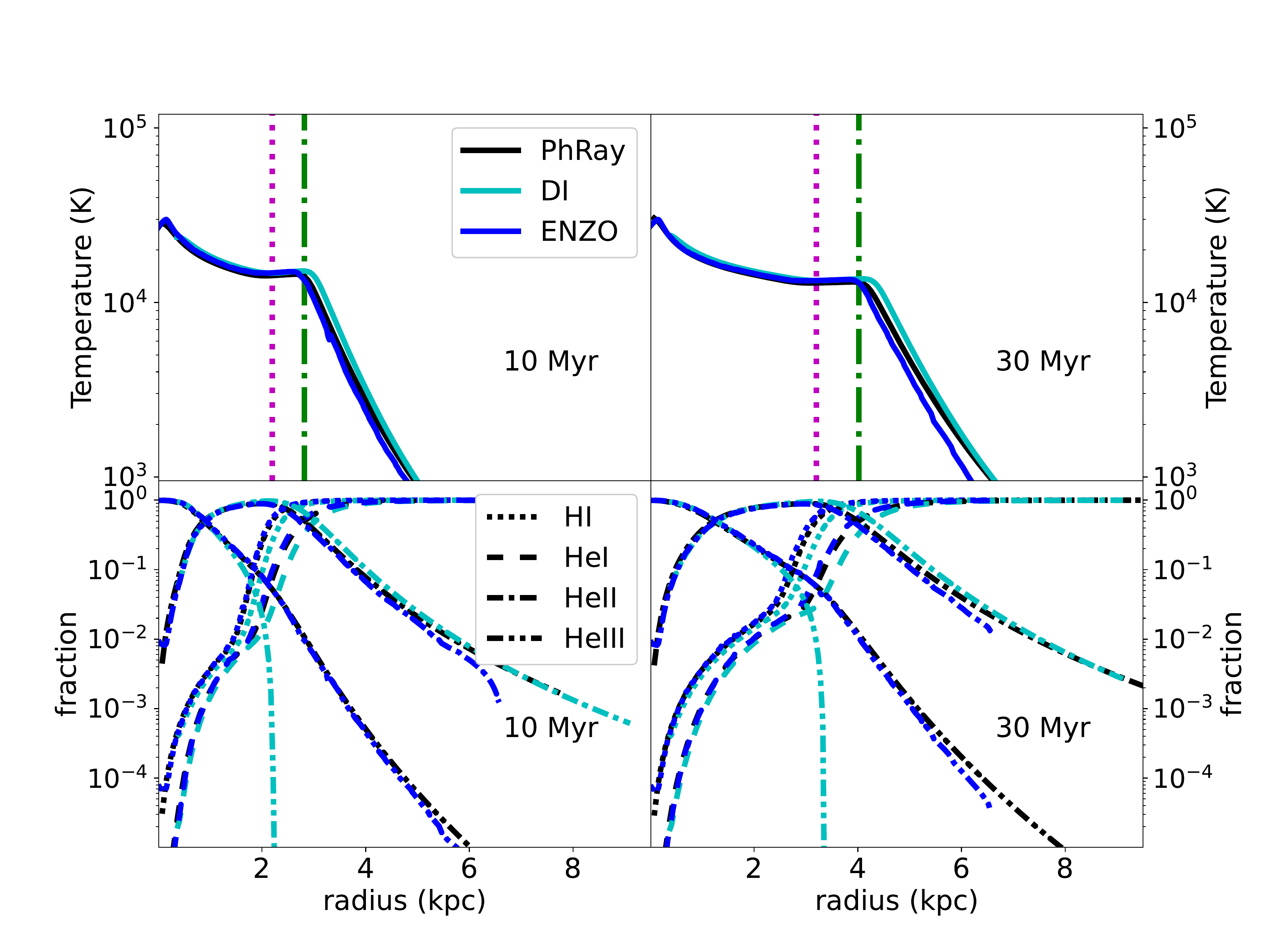}
    \caption{Reionization by a black body source with temperature $T_\mathrm{BB}=10^5$~K. Shown are results for the time-dependent photon packet code (\texttt{PhRay}, black lines), the direct integration scheme (DI, cyan lines) and \texttt{ENZO} (blue lines). Upper panels:\ Temperature profiles. The dotted (magenta) vertical line shows the \HII-front according to the \texttt{PhRay} calculation; the dot-dashed (green) vertical line shows the \HeII-front. Lower panels:\ ionization profiles for \HI\ (dotted lines), \HeI\ (dashed lines), \HeII\ (dot-dashed lines) and \HeIII (dot-dot-dashed lines).}
    \label{fig:BB_T1e5K}
\end{figure}

\begin{figure}
    \includegraphics[width=\columnwidth]{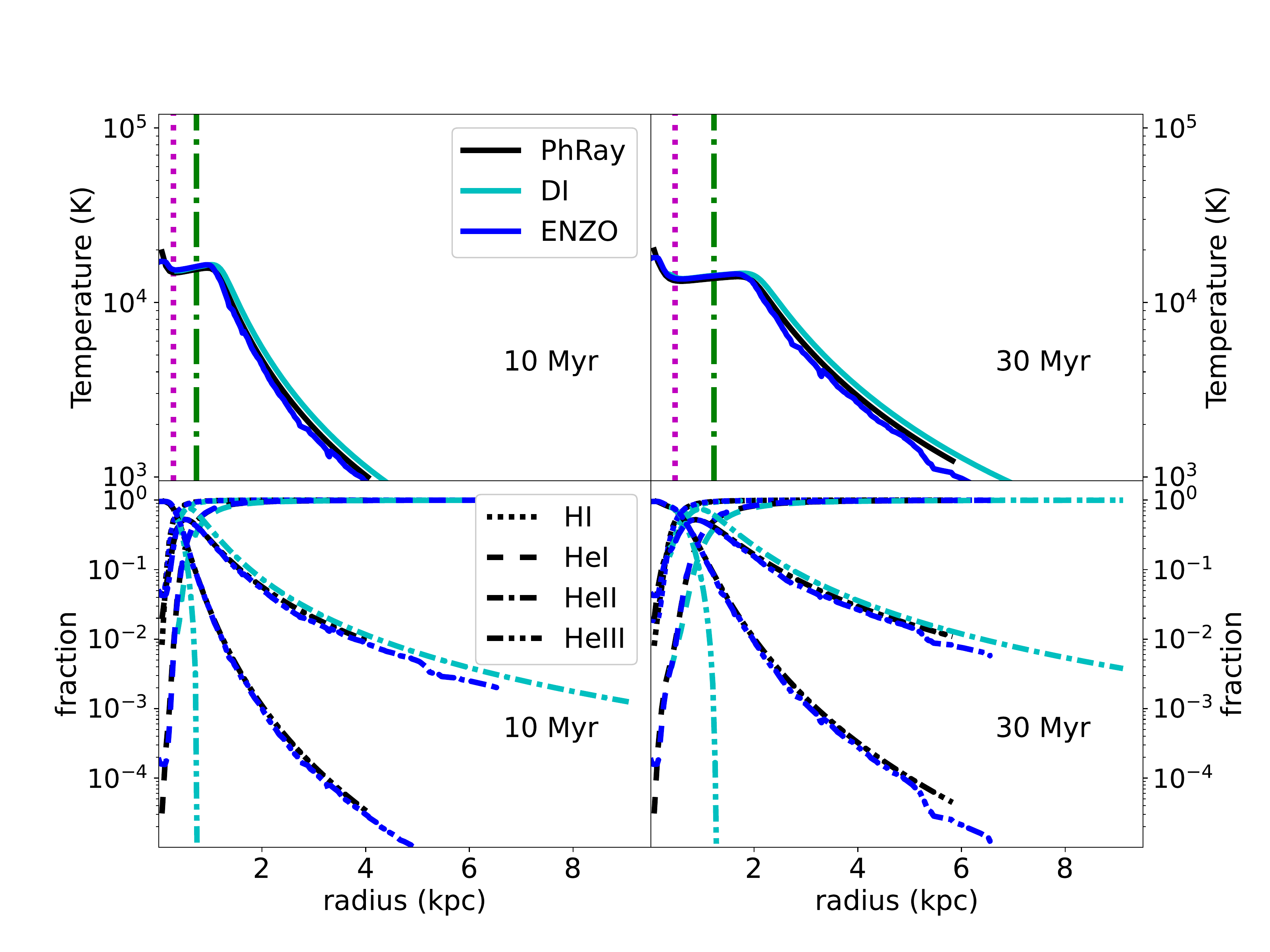}
    \caption{As in Fig.~\ref{fig:BB_T1e5K}, but for a black body source with temperature $T_\mathrm{BB}=10^6$~K.}
    \label{fig:BB_T1e6K}
\end{figure}

The temperature and ionization structure at times $t=10$~Myr and 30~Myr after the source turns on are shown in Fig.~\ref{fig:BB_T1e5K}. The temperature profiles from all the codes are in substantial agreement, although the direct integration code slightly anticipates the position of the knee in the temperature profile, where it starts its decline to $T<10^4$~K, just past the \HeII-front.

The knee in the temperature profile is reflected in the ionization profiles, for which the \HII\ and \HeII\ fronts from the direct integration scheme slightly lead the results from the photon packet codes. At distances beyond the \HeIII-front, at 0.9~kpc (1.2~kpc) at $t=10$~Myr (30~Myr), the \HeIII\ fraction from the direct integration scheme first declines gently, then decreases precipitously near 2~kpc (3~kpc), near the \HII-front at 2.2~kpc (3.2~kpc) at $t=10$~Myr (30~Myr). By contrast, both photon packet codes (\texttt{PhRay} and \texttt{ENZO}), allow leakage of \HeII-ionizing photons to larger distances sufficient to maintain partial \HeIII\ ionization. The direct integration scheme none the less produces \HeII\ fractions similar to the photon packet codes at all radii. At distances beyond 5~kpc, the \HeII\ and \HeIII\ fractions from \texttt{ENZO} decline faster than the corresponding fractions from \texttt{PhRay}. This is found to be a spatial resolution effect in the simulation volume:\ increasing the resolution in \texttt{ENZO} increases the range of agreement with the results from \texttt{PhRay}.

\subsubsection{$T=10^6$~K}
\label{subsubsec:PopIII}
The initial gas conditions are identical to those used in the $10^5$~K black body problem. The coefficient $L_{0} = 1.9\times10^{27}\,\mathrm{eV^{-2}}$
corresponds to $\dot N_{\mathrm{H}, \gamma} = 7\times10^{47}\,\mathrm{s^{-1}}$. To allow for the higher frequency peak in the Planck distribution, the upper energy bin in the \texttt{ENZO} computation is increased to
$1000\,\mathrm{eV}$ to ensure convergence on the temperature.

The temperature and ionization structure at times $t=10$~Myr and 30~Myr after the source turns on are shown in Fig.~\ref{fig:BB_T1e6K}. Whilst the fraction of emitted photons able to fully ionize helium is higher compared with the $T_\mathrm{BB}=10^5$~K black body spectrum, the results are qualitatively very similar to those for the $10^5$~K source. The exception is the position of the temperature knee, where the gas temperature starts to decline below $10^4$~K. For the $T_\mathrm{BB}=10^6$~K source, the plateau in temperature is maintained at $T\simeq(1.3-1.5)\times10^4$~K somewhat beyond the \HeII-front, with the front (shown by the dot-dashed green vertical line) positioned about half way through the temperature plateau. The temperature falls below $10^4$~K only once the \HeII\ fraction declines to below about 10 percent.

\subsection{Power-law spectra}
\label{subsec:QSO}

\subsubsection{QSO reionization}
\label{subsubsec:QSO}

We consider two reionization problems:\ (1) the reionization of the IGM at $z=6$ and (2) the reionization of the \HeII\ component of the IGM at $z=4$, both by a QSO spectrum modelled as a power law in frequency for values above the frequency $\nu_L$ of the hydrogen photoelectric threshold, $L_\nu=L_L(\nu/\nu_L)^{-\alpha_Q}$. Typical initial gas densities and ionization states are adopted at these redshifts, as explained below. The reionization is followed for times short compared with the Hubble time, so that cosmological expansion is not included:\ the gas is static. Inverse Compton cooling off the CMB is also neglected, as the characteristic cooling time is 0.5~Gyr at $z=6$ and 2~Gyr at $z=4$. The QSO spectra are modelled as $L_{S,\nu}=0.56\times10^{31}\,\mathrm{erg s^{-1} Hz^{-1}}(\nu/\nu_L)^{-0.5}$ and $L_{S,\nu}=2.0\times10^{31}\,\mathrm{erg s^{-1} Hz^{-1}}(\nu/\nu_L)^{-1.73}$. Only the photon packet codes are run for this problem since convergent results become computationally prohibitively expensive for the direct integration code.

The cut-off energies for the \texttt{ENZO} simulations for the  $L_{S,\nu}\sim\nu^{-0.5}$ and $L_{S,\nu}\sim\nu^{-1.73}$ spectra are both set to $1000\rm eV$ (see Appendix \ref{appendix:convergence} for details). The physical box size used in all the simulations is 25~Mpc, with $256^{3}$ cubic cells. For the \texttt{PhRay} spherically symmetric simulations, the grid cell size is chosen to assure the maximum initial optical depths in a single cell for \HI, \HeI\ and \HeII\ do not exceed unity.

\subsubsection{Reionization at $z=6$}
\label{subsubsec:reion}

The surrounding hydrogen density is $6.5\times10^{-5}\,\mathrm{cm}^{-3}$, corresponding to the mean IGM density at redshift $z=6$, the typical redshift when QSOs begin photoionizing the IGM. The hydrogen and helium are assumed neutral\footnote{At $z=6$, $\sim20$ percent of the volume of the IGM is expected to be neutral \citep{2022arXiv220802260G}.}, with initial gas temperature set to $T=100$~K.

\begin{figure}
    \includegraphics[width=\columnwidth]{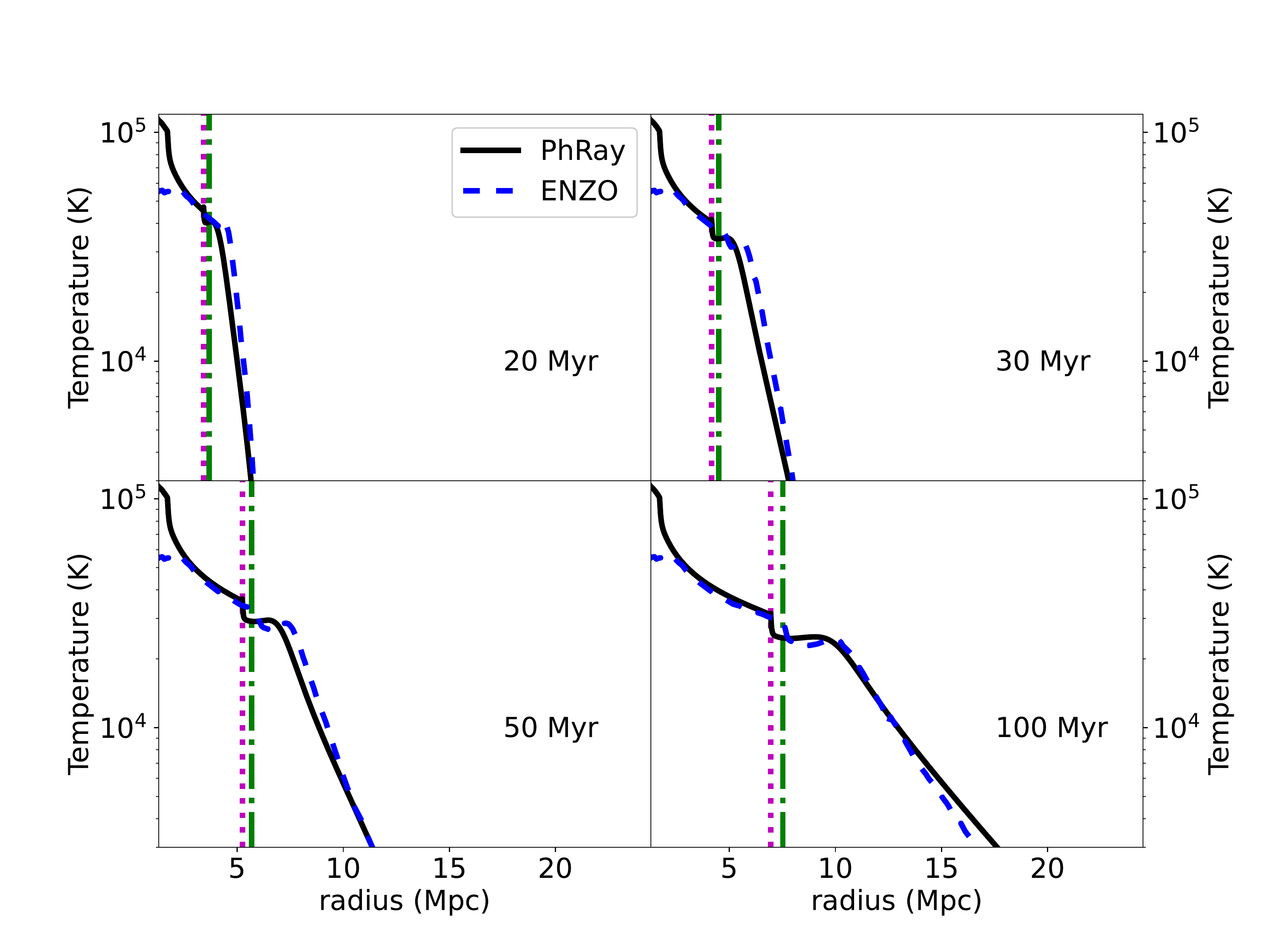}
    \includegraphics[width=\columnwidth]{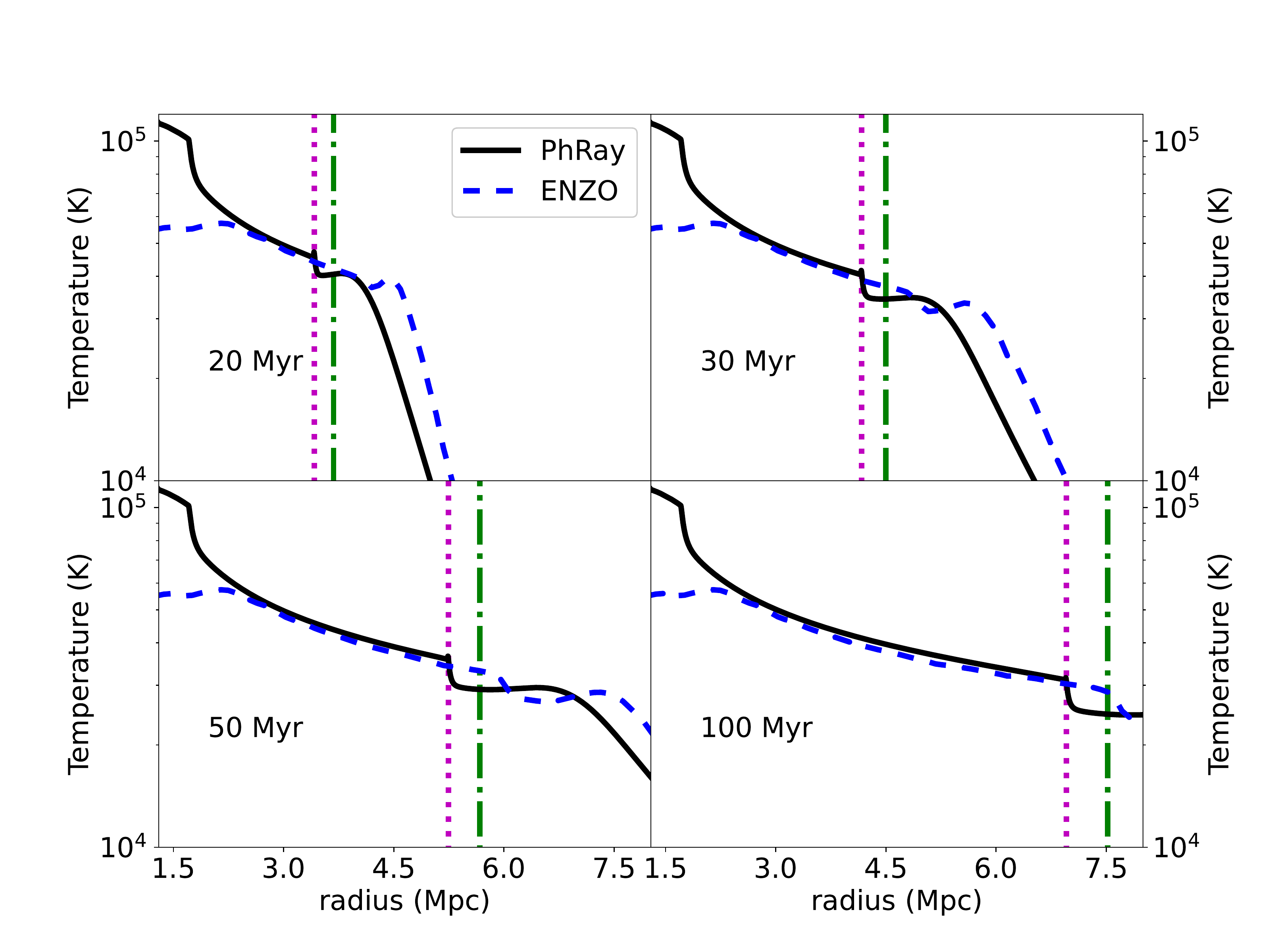}
    \caption{Temperature profiles for reionization at $z=6$ by source $L_{S,\nu}\sim\nu^{-0.5}$. The bottom set of panels shows the detailed temperature structure within the ionization fronts. Shown are the results for \texttt{PhRay} (black solid lines) and \texttt{ENZO} (dashed blue lines). The dotted magenta line in each panel shows the position of the \HII-front. The dot-dashed green line shows the leading edge of the \HeII\ zone.}
    \label{fig:Temp_z6_lnup5}
\end{figure}

\begin{figure}
    \includegraphics[width=\columnwidth]{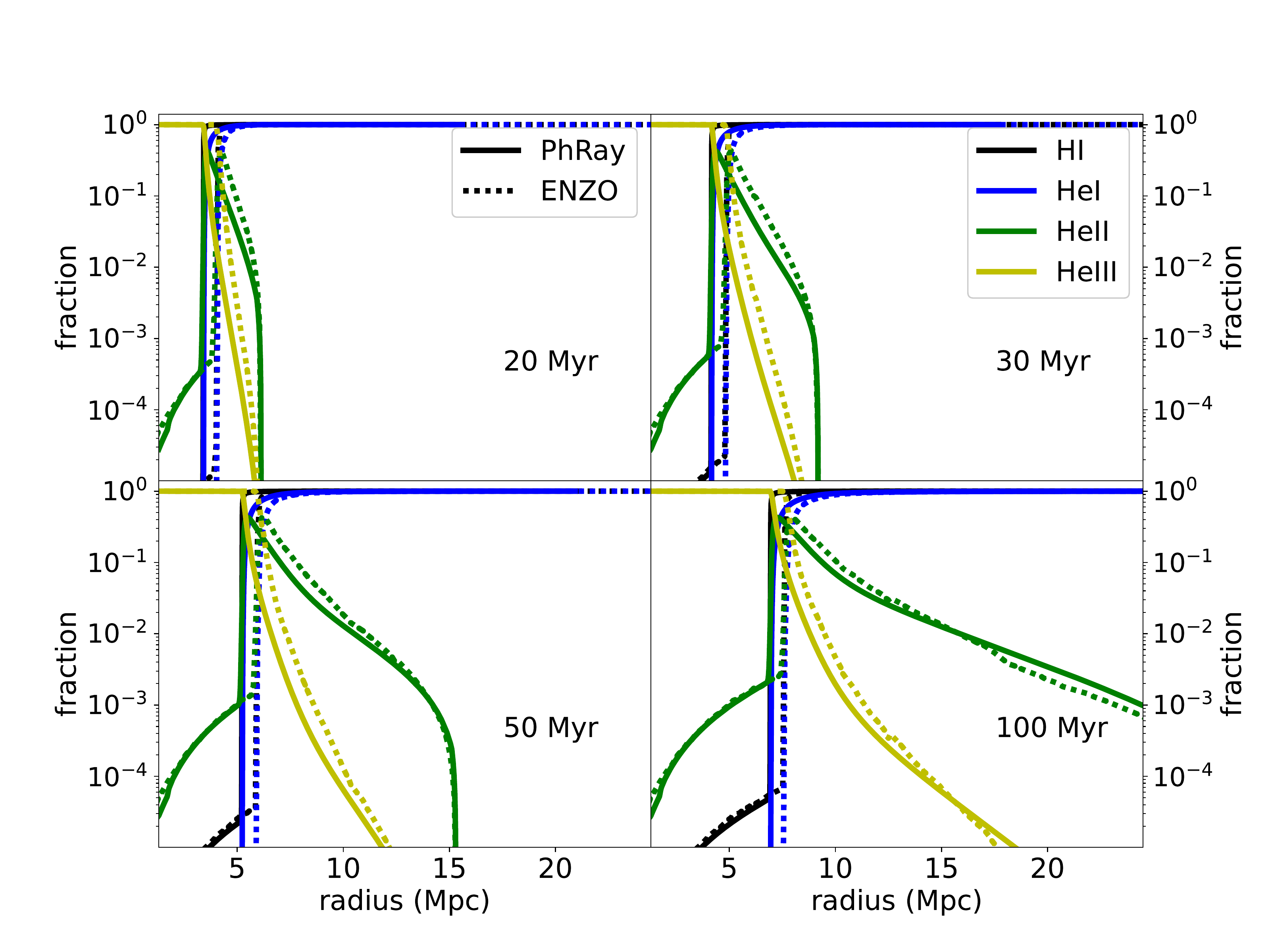}
    \caption{Ionization profiles for reionization at $z=6$ by source $L_{S,\nu}\sim\nu^{-0.5}$. Shown are the results for \texttt{PhRay} (solid lines) and \texttt{ENZO} (dotted lines), for the \HI\ (black lines), \HeI\ (blue lines), \HeII\ (green lines) and \HeIII\ (yellow lines) ionization fractions.}
    \label{fig:ion_z6_lnup5}
\end{figure}

\begin{figure}
    \includegraphics[width=\columnwidth]{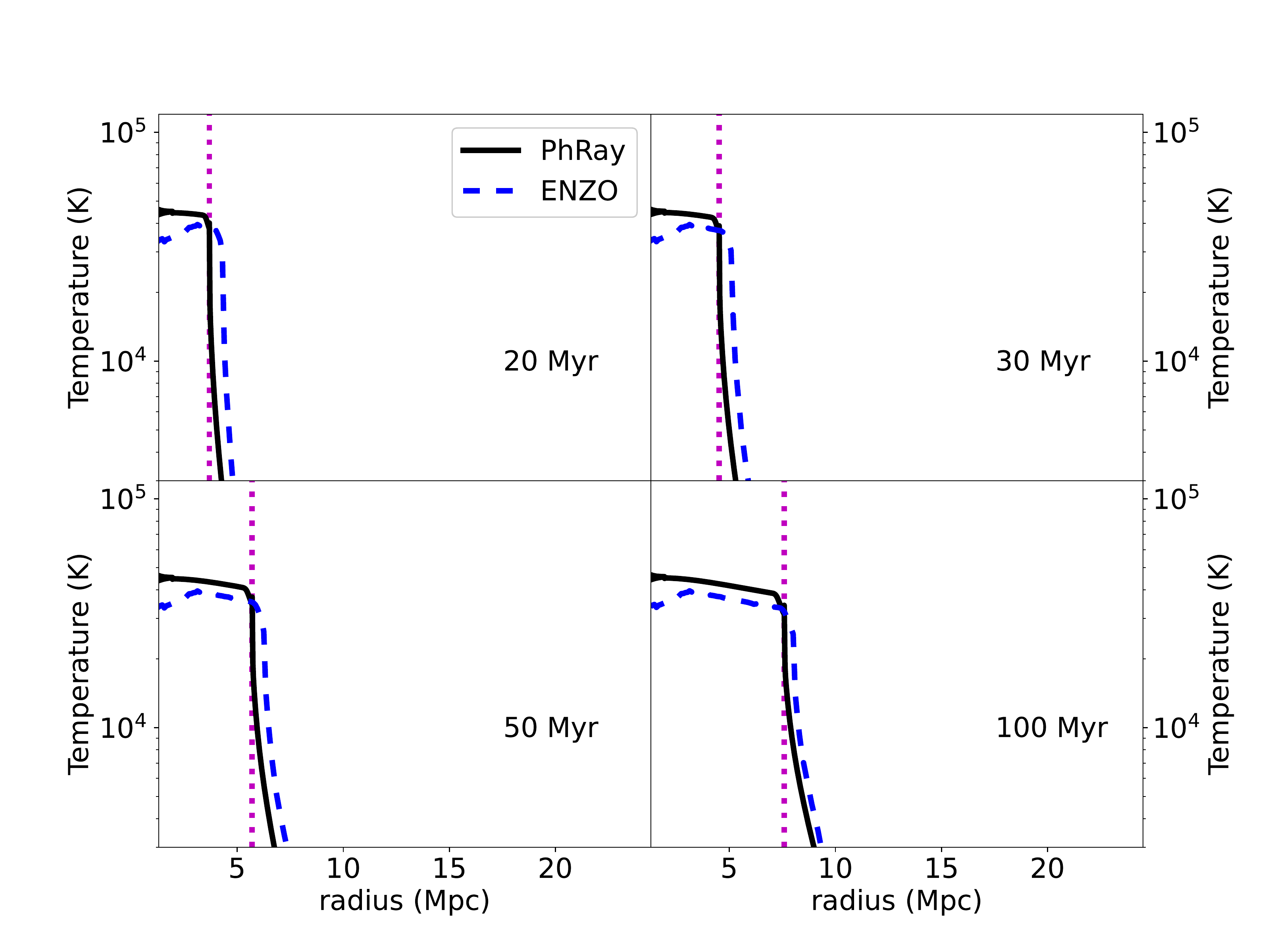}
    \caption{Temperature profiles for reionization at $z=6$ by source $L_{S,\nu}\sim\nu^{-1.73}$. Shown are the results for \texttt{PhRay} (black solid lines) and \texttt{ENZO} (dashed blue lines). The dotted magenta line shows the position of the \HII\ front.}
    \label{fig:Temp_z6_lnu1p7}
\end{figure}

\begin{figure}
    \includegraphics[width=\columnwidth]{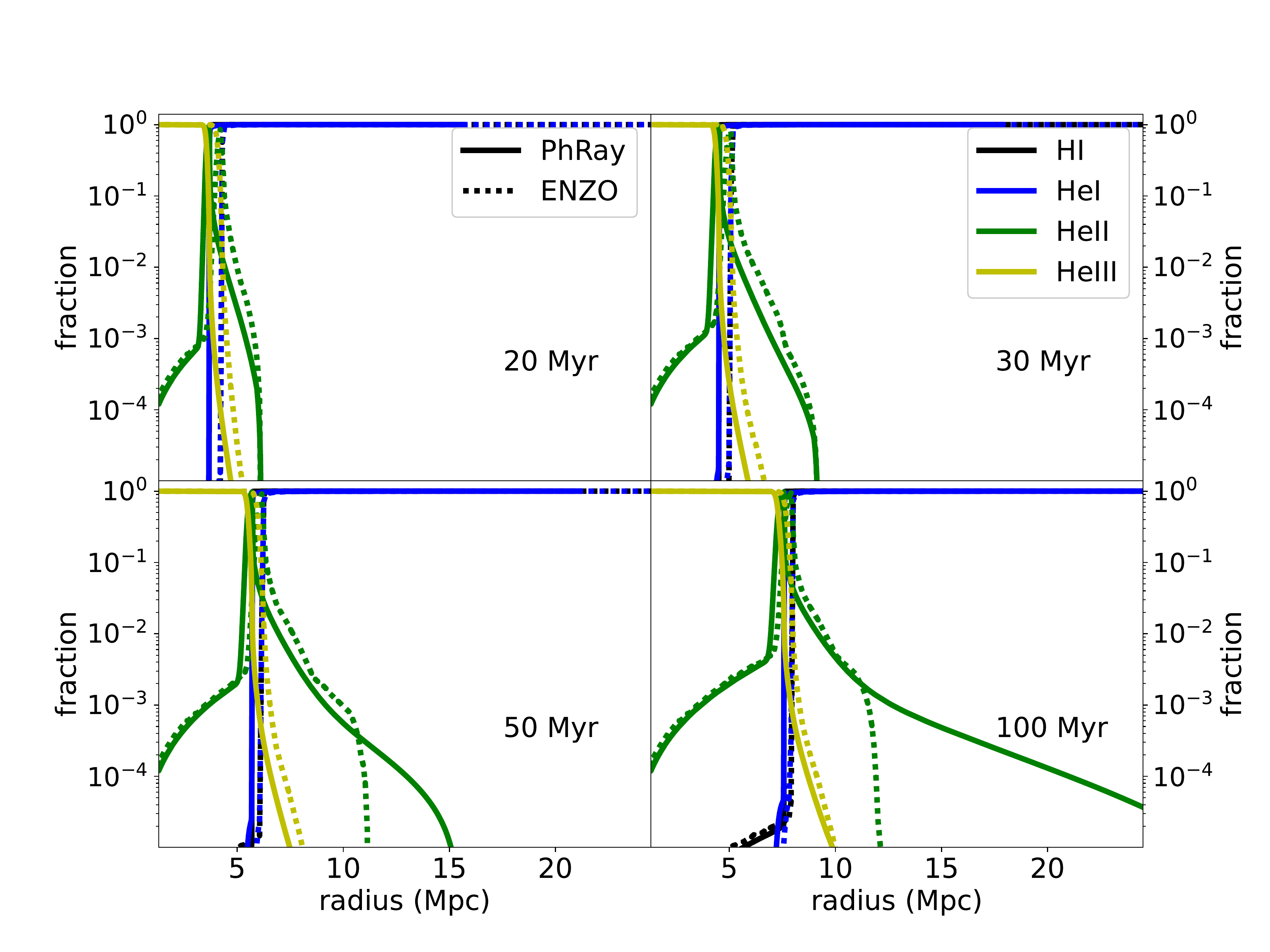}
    \caption{Ionization profiles for reionization at $z=6$ by source $L_{S,\nu}\sim\nu^{-1.73}$, as in Fig.~\ref{fig:ion_z6_lnup5}.}
    \label{fig:ion_z6_lnu1p7}
\end{figure}

Outside the central 3~Mpc, the gas temperature profiles for \texttt{PhRay} and \texttt{ENZO} substantially agree for $\alpha_Q=0.5$, as shown in Fig.~\ref{fig:Temp_z6_lnup5}. The temperature takes a sharp step down, by about $\Delta T\simeq5000$~K, at the \HII-front (shown by the dotted magenta lines). Both codes capture the temperature step as well as the temperature decline beyond the \HeII\ ionized zone. Within the inner 3~Mpc, however, the temperature from \texttt{PhRay} is boosted compared with that from \texttt{ENZO}, reaching values exceeding $10^5$~K;\footnote{We confirmed that this result is largely unaffected by inverse Compton cooling off the CMB. Including inverse Compton cooling lowers the peak temperature by only 10 percent by $t=100$~Myr.} the \HII\ region is expanding nearly at the speed of light out to this distance. This region is discussed in more detail in Sec.~\ref{sec:discussion} below.

The ionization fractions from the two codes similarly track each other closely, as shown in Fig.~\ref{fig:ion_z6_lnup5}, although the ionized hydrogen and helium regions tend to lead slightly in the \texttt{ENZO} computation. In spite of the difference in gas temperature within the central 3~Mpc, the rise in \HeII\ fractions agree well in this region. The leading edge of the \HeII-ionized region (shown in Fig.~\ref{fig:Temp_z6_lnup5} by the dot-dashed green line) extends slightly beyond the \HII\ region (shown by the dotted magenta line). Comparison with the $T_\mathrm{BB}=10^6$~K black body spectrum case in Fig.~\ref{fig:BB_T1e6K}, for which the \HII\ end \HeII\ fronts are more clearly separated, shows that the ledge in high temperature actually extends beyond the \HeII-front, into the region of partial \HeII\ ionization. For the power-law spectrum case here, the temperature falls below $2\times10^4$~K only for a \HeII\ fraction below 5--7 percent. At $t=100$~Myr, a low level of \HeII\ ionization persists to the edge of the simulation volume, with ionization fraction $x_{HeII}>5\times10^{-4}$ and $T>350$~K, large compared with the initial temperature of 100~K.

For the $\alpha_Q=1.73$ spectrum, as shown in Fig.~\ref{fig:Temp_z6_lnu1p7}, the boost in temperature in the central 3~Mpcs for \texttt{PhRay} compared with \texttt{ENZO} is smaller than for the $\alpha_Q=0.5$ source. The gas temperature declines abruptly at the \HII\ front, shown by the dotted magenta lines in Fig.~\ref{fig:Temp_z6_lnu1p7}. The leading edge of the \HeII\ region almost exactly tracks the \HII\ front (with positions agreeing to better than 1 percent) at all times, with no ledge in high temperature extending beyond as in the $\alpha_Q=0.5$ spectrum case.

The ionized regions from \texttt{ENZO} again slightly lead those from \texttt{PhRay}, as shown in Fig.~\ref{fig:ion_z6_lnu1p7}. The region of low \HeII\ ionization extends further for the \texttt{PhRay} calculation than for \texttt{ENZO}; the extent is limited by the higher energy photon cut-off in \texttt{ENZO}. At $t=100$~Myr, $x_{HeII}>1.8\times10^{-5}$ out to the edge of the \texttt{PhRay} simulation volume of 26~Mpc radius, with $T>108$~K. Compared with the initial temperature of 100~K, the amount of heating is small at these radii.

\subsubsection{Reionization of \HeII\ at $z=4$}
\label{subsubsec:HeIIreion}

The surrounding hydrogen density is $2.4\times10^{-5}\,\mathrm{cm}^{-3}$, corresponding to the mean IGM density at redshift $z=4$, the typical redshift when QSOs begin photoionizing \HeII\ in the IGM. The initial hydrogen neutral fraction is set at $x_\mathrm{HI}=2\times10^{-5}$ and the \HeI\ and \HeIII\ helium fractions $x_\mathrm{HeI}=9\times10^{-6}$ and $x_\mathrm{HeIII}=0$. These correspond approximately to the ionization levels for the ultra-violet (UV) metagalactic background at $z=4$ \citep{2012ApJ...746..125H} in a region for which \HeII\ has not yet been ionized. The initial gas temperature is set to $T=10^4$~K.

\begin{figure}
    \includegraphics[width=\columnwidth]{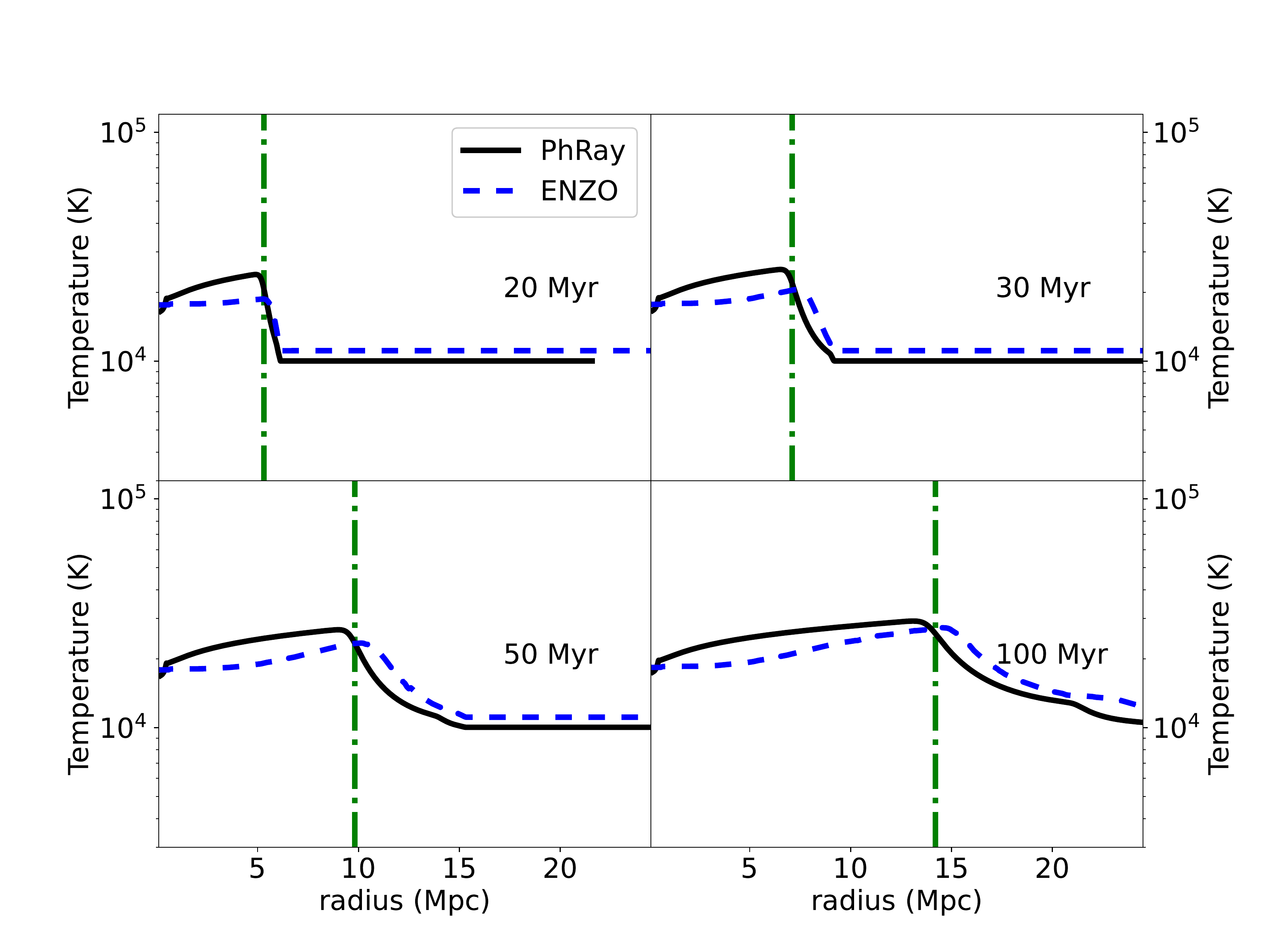}
    \caption{Temperature profiles following \HeII\ reionization at $z=4$ by a source $L_{S,\nu}\sim\nu^{-0.5}$. Shown are the results for \texttt{PhRay} (black solid lines) and \texttt{ENZO} (dashed blue lines). The dot-dashed green line in each panel shows the position of the \HeIII-front.}
    \label{fig:Temp_z4_lnup5}
\end{figure}

\begin{figure}
    \includegraphics[width=\columnwidth]{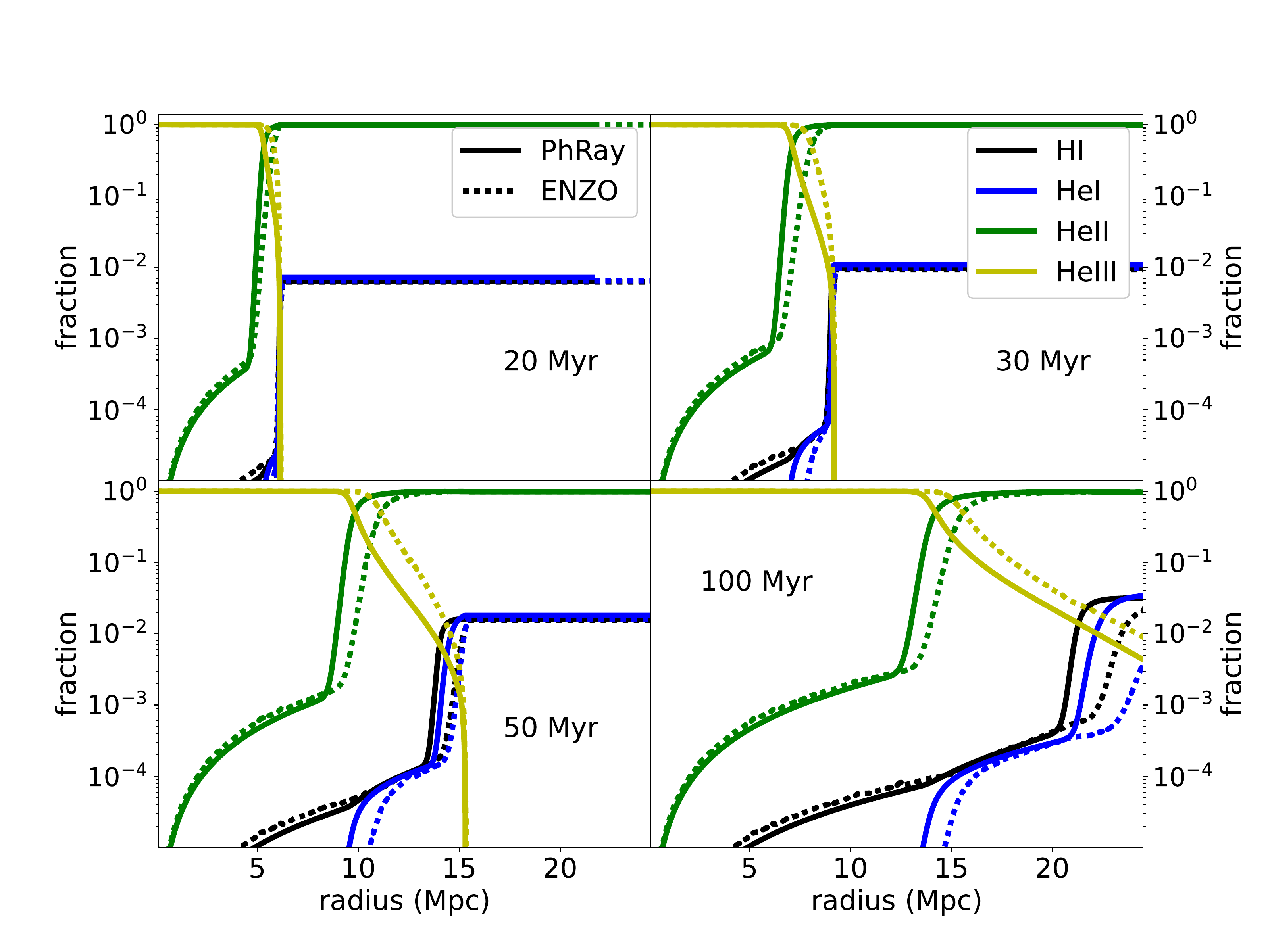}
    \caption{Ionization profiles following \HeII\ reionization at $z=4$ by a source $L_{S,\nu}\sim\nu^{-0.5}$, as in Fig.~\ref{fig:ion_z6_lnup5}.}
    \label{fig:ion_z4_lnup5}
\end{figure}

\begin{figure}
    \includegraphics[width=\columnwidth]{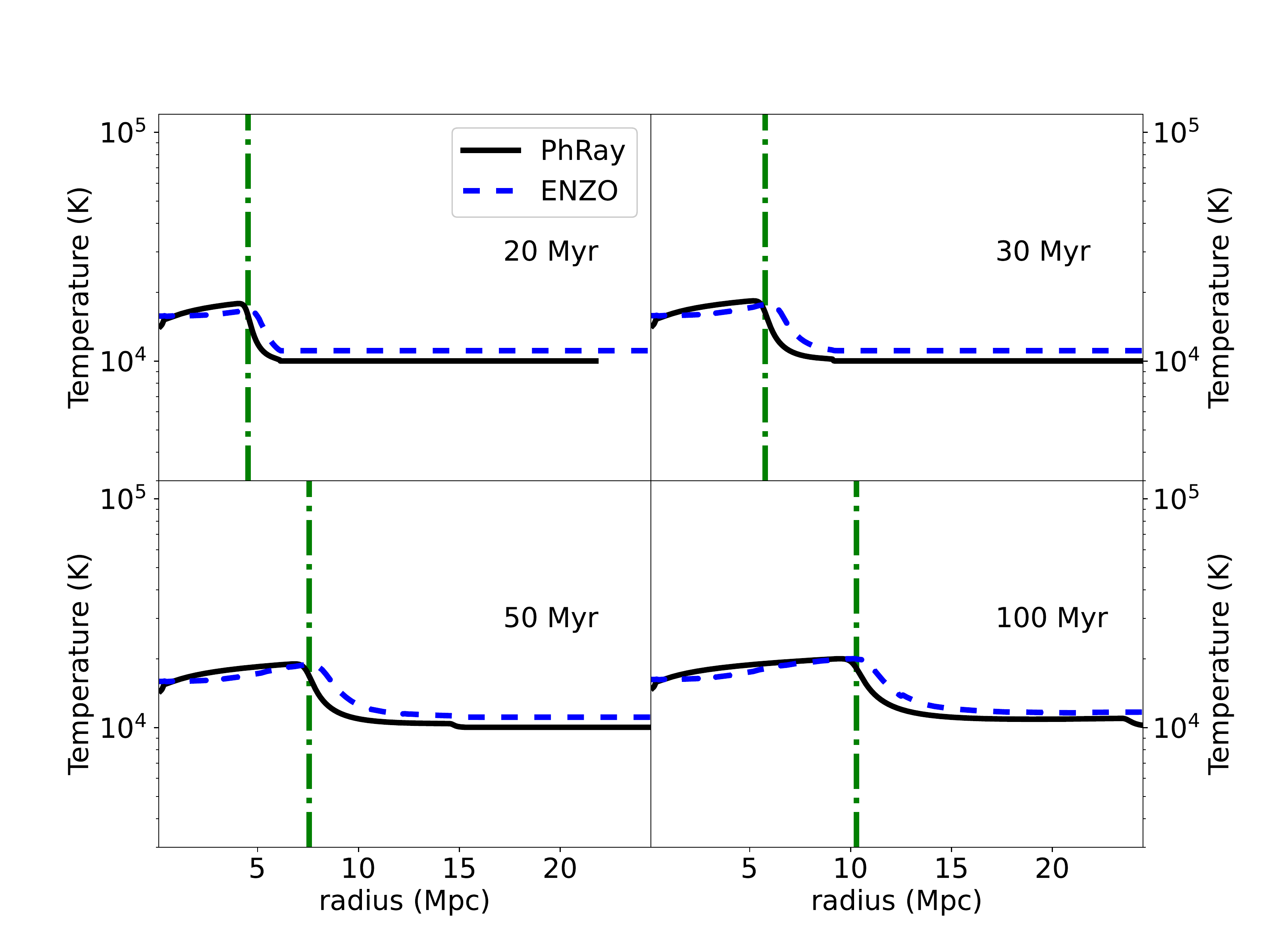}
    \caption{Temperature profiles following \HeII\ reionization at $z=4$ by a source $L_{S,\nu}\sim\nu^{-1.73}$. Shown are the results for \texttt{PhRay} (black solid lines) and \texttt{ENZO} (dashed blue lines). The dot-dashed green line shows the position of the \HeIII-front.}
    \label{fig:Temp_z4_lnu1p7}
\end{figure}

\begin{figure}
    \includegraphics[width=\columnwidth]{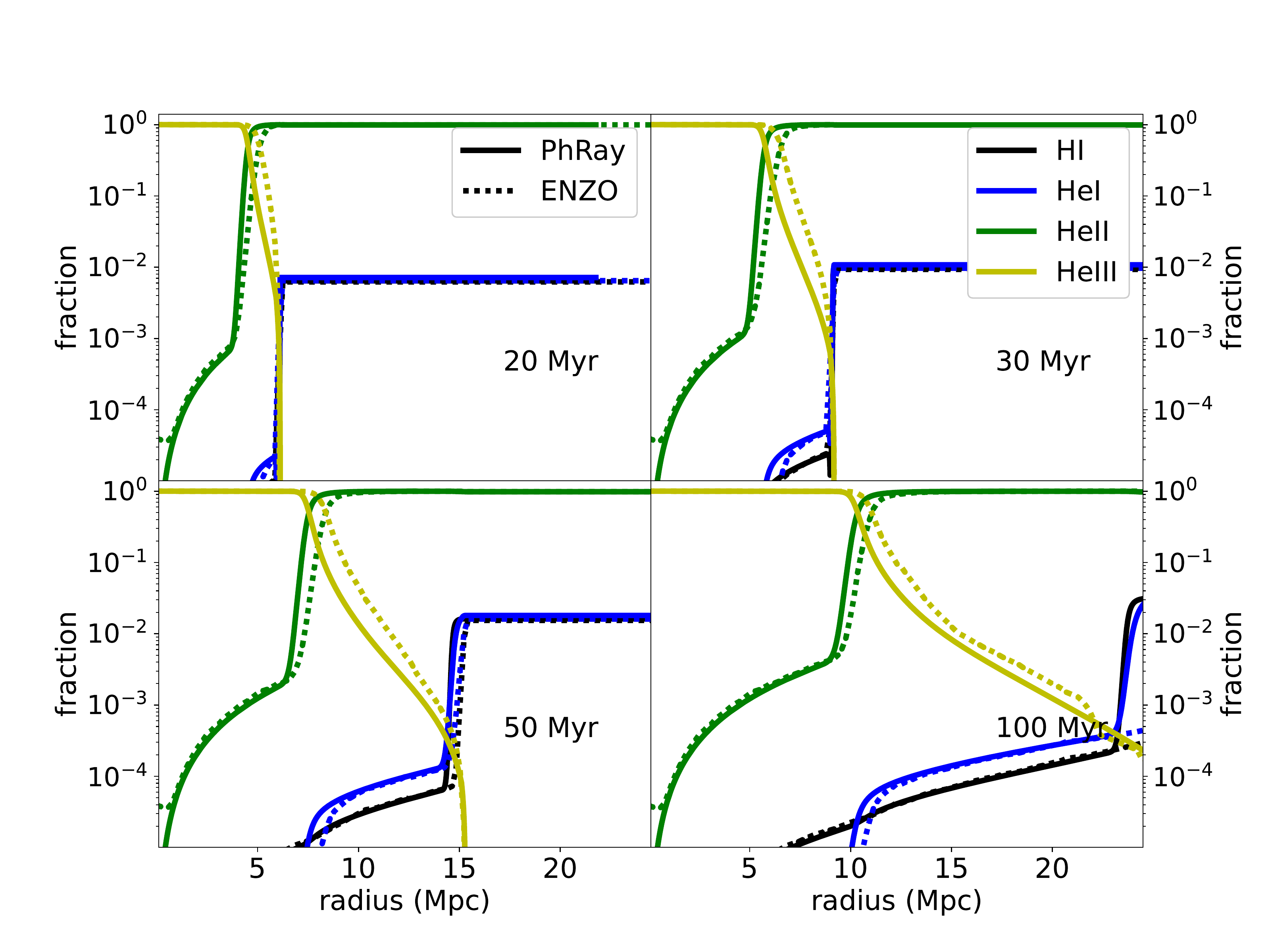}
    \caption{Ionization profiles following \HeII\ reionization at $z=4$ by a source $L_{S,\nu}\sim\nu^{-1.73}$, as in Fig.~\ref{fig:ion_z6_lnup5}.}
    \label{fig:ion_z4_lnu1p7}
\end{figure}

For the $\alpha_Q=0.5$ spectrum, as shown in Fig.~\ref{fig:Temp_z4_lnup5}, the gas temperature is elevated behind the \HeIII-front relative to the temperature of the ambient gas. Whilst the \texttt{PhRay} and \texttt{ENZO} ionization levels agree well within the \HeIII\ region, as shown in Fig.~\ref{fig:ion_z4_lnup5}, the \texttt{PhRay} temperature somewhat exceeds that of \texttt{ENZO} by about 4000~K. As discussed below, this is a consequence of near luminal expansion of the \HeIII-front once the QSO turns on. Ahead of the \HeIII-front, the temperatures are in good agreement, although the \texttt{ENZO} temperature is slightly higher than the \texttt{PhRay} temperature.

As for the $z=6$ simulations, the \texttt{ENZO} ionization regions slightly lead those from \texttt{PhRay} (Fig.~\ref{fig:ion_z4_lnup5}), with the more ionized \HII\ and \HeII\ regions expanding somewhat more rapidly for \texttt{ENZO}. Otherwise the ionization fractions are in good agreement outside the \HeIII\ region. At distances from the source beyond the light front, the ionization fractions remain constant with distance, reflecting the initial conditions. Because there is no ambient UV photoionizing background field, the ionization level at these distances is evolving as hydrogen and helium gradually recombine. 

The temperature is again elevated out to the \HeIII-front for the $\alpha_Q=1.73$ spectrum relative to the ambient gas temperature, as shown in Fig.~\ref{fig:Temp_z4_lnu1p7}, but not by as much as for the $\alpha_Q=0.5$ spectrum. The temperatures from \texttt{PhRay} and \texttt{ENZO} agree well, although the \texttt{ENZO} temperature slightly exceeds that of \texttt{PhRay} beyond the \HeIII-front. This is consistent with a slightly faster expansion of the \HeIII-front from \texttt{ENZO} compared with \texttt{PhRay}, as shown in Fig.~\ref{fig:ion_z4_lnu1p7}.

\section{Discussion}
\label{sec:discussion}

Both the direct integration and photon packet codes recover the principal ionized zones of hydrogen and helium produced by the black-body and power-law spectral sources. Several discrepancies, however, are found. We discuss the differences that are particularly pertinent to measurements of the IGM. We focus on differences in the near zones, where hydrogen and helium are nearly fully ionized, and the far zones, where the hydrogen and helium are nearly neutral.

\subsection{Near zone}
\label{subsec:nearzone}

\subsubsection{Black-body spectra}
\label{subsubsec:nzStars}

For the $T_\mathrm{BB}=10^5$~K black-body spectrum, the hydrogen-ionizing photon emission rate $\dot N_{\mathrm{H}, \gamma}=5\times10^{48}\,\mathrm{s}^{-1}$ corresponds to an expansion rate of the \HII-front, before radiative recombinations become important, given by balancing the emission rate to the rate at which hydrogen atoms are ionized:

\begin{equation}
    r_\mathrm{HII} = \left(\frac{3}{4\pi}\frac{\dot N_{\mathrm{H}, \gamma}t}{ n_\mathrm{H}}\right)^{1/3}\simeq 1.2 t_\mathrm{Myr}^{1/3}\,\mathrm{kpc},
    \label{eq:rhii}
\end{equation}
where $t_\mathrm{Myr}$ is the time since the source turned on in units of $10^6$~yr and a hydrogen density $n_\mathrm{H}=0.76\times10^{-3}\,\mathrm{cm}^{-3}$ has been assumed\footnote{Eq.~(\ref{eq:rhii}) is an approximation assuming all ionizing photons are absorbed at the ionization front. In practice, sufficiently high energy photons continue un-absorbed because of their long mean free paths, but they make up only a small fraction of all the photons.}. For the $T_\mathrm{BB}=10^6$~K black-body spectrum, with the lower hydrogen-ionizing photon emission rate $\dot N_{\mathrm{H}, \gamma}=7\times10^{47}\,\mathrm{s}^{-1}$, the expansion rate is about half as fast, $r_\mathrm{HII}\simeq 0.6t_\mathrm{Myr}^{1/3}\,\mathrm{kpc}$.

The growth of the \HII\ region for the $T_\mathrm{BB}=10^5$~K source agrees well with the theoretical expectation, with the \HII-front (defined at the position where $x_\mathrm{HII}=0.5$), occurring within 15 percent of the prediction of Eq.~(\ref{eq:rhii}), although falling systematically slightly short. The agreement is poorer for the harder $T_\mathrm{BB}=10^6$~K spectrum, with the \HII-front lagging far behind the prediction. The discrepancies may be attributed to the presence of helium. For $T_\mathrm{BB}=10^5$~K, about half the hydrogen-ionizing photons may ionize helium. After subtracting these, the predicted position of the \HII-front decreases by about 20 percent. For $T_\mathrm{BB}=10^6$~K, 99 percent of the hydrogen-ionizing photons may also ionize helium. Removing these decreases the predicted radius of the \HII-front by about a factor of 5, in good agreement with the computations. The inclusion of helium thus requires accounting for the sharing of photons that may ionize more than a single species, which will depend on the relative abundances of the species in general, as well as on their relative cross sections.

The ionization structures for the black-body spectra also agree well between the codes. Nearly perfect agreement is found for the photon packet codes \texttt{PhRay} and \texttt{ENZO}. The ionization fronts from the direct integration scheme, however, slightly lead the positions from the photon packet codes.

\subsubsection{Power-law spectra}
\label{subsubsec:nzQSO}

\begin{figure}
    \includegraphics[width=\columnwidth]{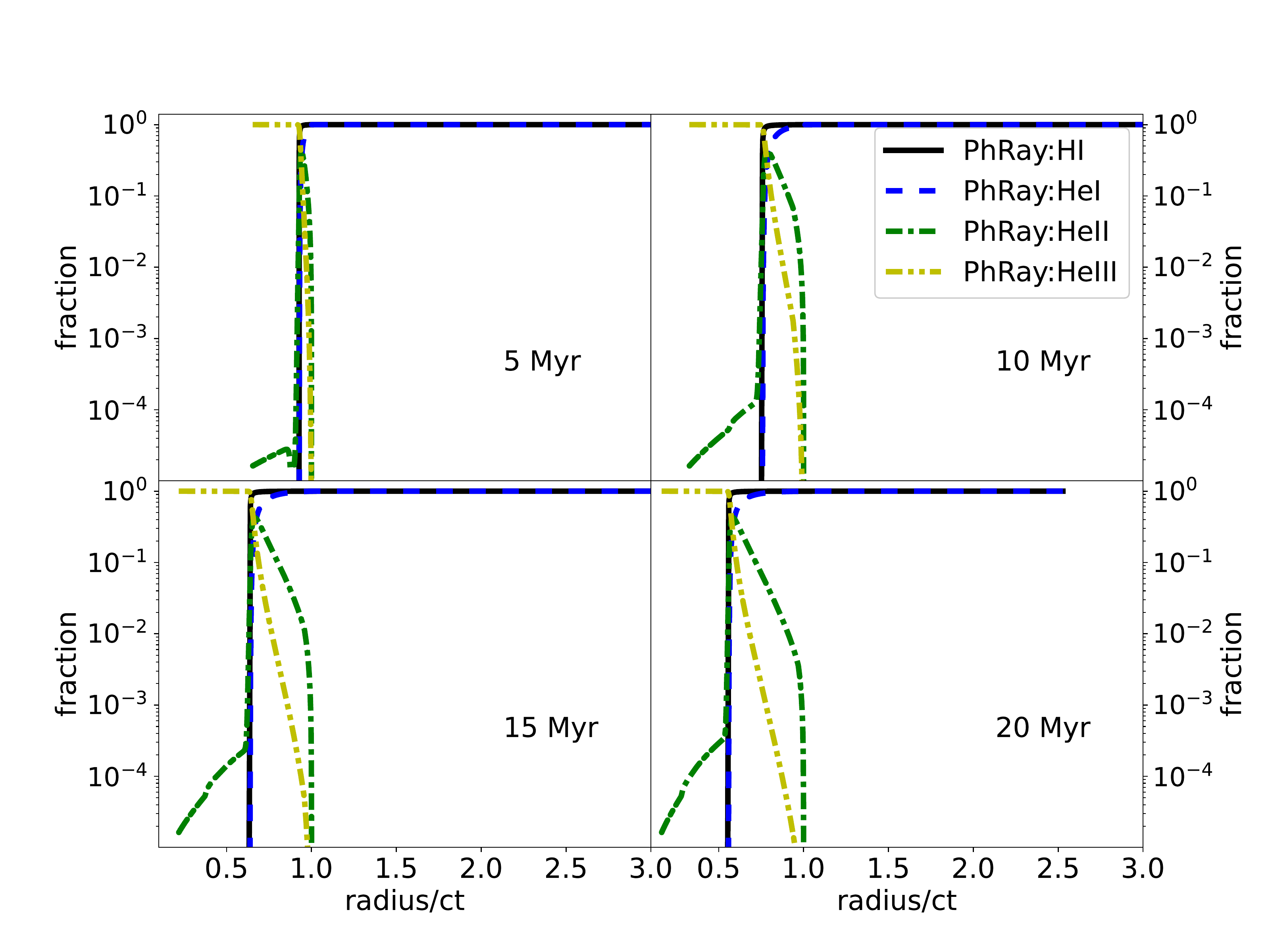}
    \caption{Ionization profiles shown vs the light front distance for reionization at $z=6$ by a source $L_{S,\nu}\sim\nu^{-0.5}$, from solving the full time-dependent RT equation.}
    \label{fig:ion_z6_ct_lnup5}
\end{figure}

\begin{figure}
    \includegraphics[width=\columnwidth]{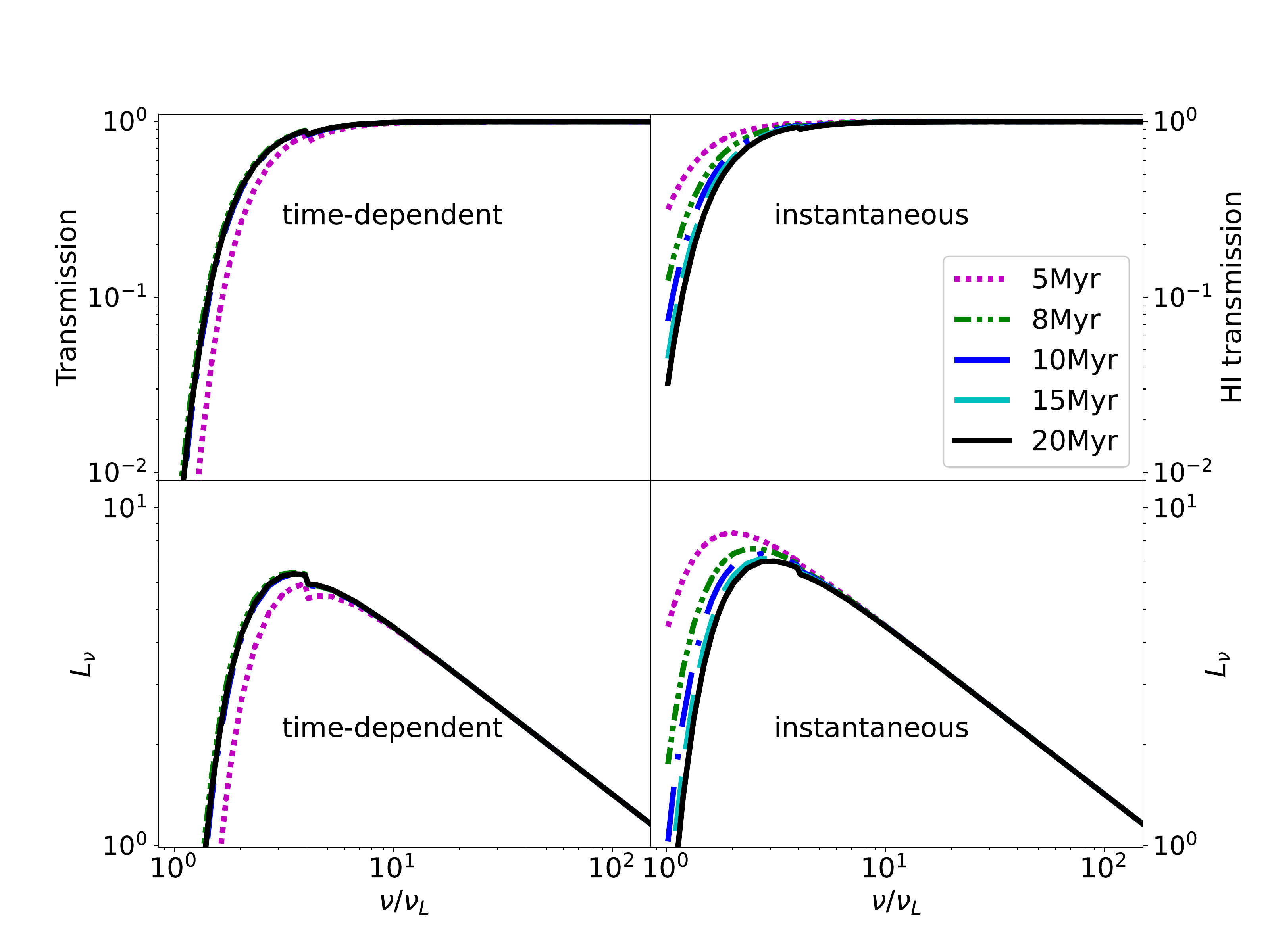}
    \caption{Comparison of the transmitted radiation for reionization at $z=6$ by a source $L_{S,\nu}\sim\nu^{-0.5}$, for the solution to the full time-dependent RT equation (left panels) and in the instantaneous (ISLA) limit (right panels). Upper panels:\ Transmission profiles $e^{-\tau_\nu}$ at \HII-front. Lower panels:\ Transmitted luminosity $L_\nu=L_{S,\nu}e^{-\tau_\nu}$ at \HII-front.} 
    \label{fig:trans_z6_ct_lnup5}
\end{figure}

Outside the inner 3~Mpc, but still within the highly ionized regions, the temperatures found by \texttt{PhRay} and \texttt{ENZO} for the test problems for IGM conditions at $z=6$ agree well. This is a significant achievement of the probabilistic formulation of the radiative transfer problem, as the initial optical depth per cell in the \texttt{ENZO} computation is 124, compared with an optical depth of unity in \texttt{PhRay}. The sharpness of the \HII-front permits a generous optical depth criterion, making the problem practical. By contrast, the convergence requirements that the optical depth per zone not exceed unity, along with a higher number of frequency bins, renders a direct integration of the radiative transfer equation computationally impractical.

For the test problem with IGM conditions at $z=4$, the temperatures between \texttt{PhRay} and \texttt{ENZO} agree well, although the temperature from \texttt{PhRay} somewhat exceeds that of \texttt{ENZO} within the \HeIII\ region. The ionization fronts from \texttt{ENZO} also slightly lead those from \texttt{PhRay}. This could over-estimate the size of the expected \HeIII\ zone predicted for a given QSO spectrum and age. Agreement improves on increasing the spatial resolution for \texttt{ENZO} from $128^3$ to $256^3$ zones, corresponding to decreasing the \HeII\ optical depth at the photoelectric threshold from 1.9 to 0.9.

The higher temperature found by \texttt{PhRay} compared with \texttt{ENZO} within the inner 3~Mpc for IGM conditions at $z=6$, and for IGM conditions at $z=4$ out to the \HeIII-front, especially for the $\alpha_Q=0.5$ spectrum, is a consequence of the rapid expansion of the ionized region around the source. In the test problem for $z=6$, the \HII\ region expands nearly at the velocity of light, whilst in the $z=4$ problem, the \HeIII\ region has near luminal expansion. This may be seen by equating the output rate of ionizing photons to the rate at which gas is photoionized, when recombinations may be neglected. The criterion that the expansion of the zone becomes subluminal is then

\begin{equation}
    \frac{1}{4\pi r_i^2}\frac{\dot N_\gamma}{n} < c,
    \label{eq:sublum}
\end{equation}
where $\dot N_\gamma$ is the production rate of ionizing photons and $n$ is the density of the species being ionized.\footnote{\citet{2003AJ....126....1W} give for the evolution of an I-front radius $R_I$, allowing for the finite travel time of light, $\dot N_\gamma(t-R_I/c)=(4\pi/3)R_I^3n$, where $n$ is the density of the species being photoionized. This corresponds to an expansion velocity $dR_I/dt = c\dot N_\gamma/(\dot N_\gamma + 4\pi R_I^2nc)$.} The criterion in Eq.~(\ref{eq:sublum}) then corresponds to an I-front velocity of 0.5c. The production rate of all photons above the \HI\ threshold for $\alpha_Q=0.5$ and 1.73 is
$1.7\times10^{57}\,\mathrm{s}^{-1}$. For a
hydrogen density $6.5\times10^{-5}\,\mathrm{cm}^{-3}$, the \HII\ region expansion will become sub-luminal only at $r\gsim 2.7$~Mpc, or somewhat smaller allowing for some photons to ionize helium. This is consistent with the agreement in the temperatures between \texttt{PhRay} and \texttt{ENZO} at $r>2.5$~Mpc, where they both give $T\simeq5.5\times10^4$~K for $\alpha_Q=0.5$ and $T\simeq4\times10^4$~K for $\alpha_Q=1.73$, as shown in Figs.~\ref{fig:Temp_z6_lnup5} and \ref{fig:Temp_z6_lnu1p7}.

Similarly, as shown in Figs.~\ref{fig:Temp_z4_lnup5} and \ref{fig:Temp_z4_lnu1p7}, enhancements in temperature are found at $z=4$ where the hydrogen has already been ionised (and the helium singly ionised). (The enhancement is small for the softer $\alpha_Q=1.73$ spectrum.) In this case, the boost in temperature results from the rapidly expanding \HeIII-fronts. From Eq.~(\ref{eq:sublum}), taking $n=n_\mathrm{He}$ and considering \HeII-ionizing photons, the expansion of the \HeIII\ region becomes sub-luminal only for $r>11$~Mpc for $\alpha_Q=0.5$ and $r>5$~Mpc for $\alpha_Q=1.73$. The temperature enhancement is confined to the region with substantial \HeIII\ ionization ($x_\mathrm{HeIII}>0.5$), shown by the magenta lines in Figs.~\ref{fig:Temp_z4_lnup5} and \ref{fig:Temp_z4_lnu1p7}. The boost persists until the region of $x_\mathrm{HeIII}>0.5$ reaches the luminal expansion limiting radius.

The reason near-luminal expansion of an ionization region gives rise to a boost in temperature is illustrated in Figs.~\ref{fig:ion_z6_ct_lnup5} and \ref{fig:trans_z6_ct_lnup5} for the $\alpha_Q=0.5$ spectrum at $z=6$. Fig.~\ref{fig:ion_z6_ct_lnup5} shows the ionization fractions as a function of position in units of the light front ($r=ct$) from the time-dependent code \texttt{PhRay}, which tracks all photon packets since they were emitted until they are absorbed. The time for the ionization fronts to reach $\sim2.5$~Mpc and become sub-luminal is 8~Myr. The ionization fronts then begin to slip increasingly behind the light front.

As long as the ionization fronts keep up with the light front, the gas encountered by the photons is largely neutral. As a consequence, the lower energy photons are rapidly absorbed by the gas. Most of the photoionization is carried out by the surviving most energetic photons. Once the ionization front becomes sub-luminal, the photon packets that arrive at the front include proportionately more lower energy photons from the source, and the amount of energy deposited in the gas per ionization decreases. This is shown in the left panels of Fig.~\ref{fig:trans_z6_ct_lnup5}. The median energy at which photons are transmitted is higher at times $t<8$~Myr, with the peak in the transmitted luminosity $L_\nu$ shifted towards higher energies. By $t>8$~Myr, the photon luminosity profile at the \HII-front reflects the transmission through the intervening ionization structure.

By contrast, rather than tracking photon packets since they were emitted, \texttt{ENZO} recasts new rays at each time step and computes the instantaneous radiative transfer along the rays with a new set of photon packets launched from the source. At $t=5$~Myr, the intervening gas between the source and the \HII-front removes fewer low energy photons (above the ionization threshold energy) than would have been removed from photon packets that were moving only very slightly ahead of the ionization front, as in the \texttt{PhRay} computation. This is shown in the right panels of Fig.~\ref{fig:trans_z6_ct_lnup5}. The transmitted luminosity $L_\nu$ at the \HII-front peaks at a lower energy compared with the time-dependent computation in the left panel. By $t>15$~Myr, the transmission factor and $L_\nu$ have relaxed to those for the time-dependent RT equation solution once the \HII-front has become sub-luminal. Thereafter, the temperatures from the time-dependent (\texttt{PhRay}) and instantaneous (\texttt{ENZO}) computations agree. The gas heated earlier, during the luminal expansion phase, in the time-dependent RT equation solution from \texttt{PhRay} at $r<2.5$~Mpc, however, remains hotter compared with the temperature computed in the instantaneous (ISLA) limit by \texttt{ENZO} because of the long cooling time.

The difference in temperature in the near-zone has possible implications for metagalactic UV background (UVBG) or QSO lifetime estimates from proximity zone measurements, as these depend on the \HI\ or \HeII\ fraction in the vicinity of QSOs. Models based on instantaneous photo-ionization may underestimate the gas temperature, and so overestimate the recombination rate and \HI\ or \HeII\ fraction. This may result in an under-estimate in the size of the proximity zone around a QSO for a given UVBG level or QSO age, and so to an under-estimate of the UVBG level or over-estimate of the QSO age needed to agree with the proximity zone measured. The additional near zone heating may also boost the \Lya\ photon emission rate through collisional excitation of \HI\ in the ionization front during the luminal expansion phase. The increase in temperature may also affect the \Lya\ forest power spectrum at wavenumbers $k\gsim0.003\,\mathrm{s\,km^{-1}}$, corresponding to the sizes of the luminal expansion regions.

The discrepancy in the predictions for the near zone between the time-dependent and ISLA solutions to the radiative transfer equation when ionization fronts expand near the speed of light poses a dilemma for photon packet radiative transfer codes. Solving the time dependent radiative transfer equation requires assigning a finite velocity to the photon packets and retaining all photon packets emitted during any previous time step until they exit the grid. This imposes an impractical memory demand on the computations. The correct size of the ionization regions may instead  be computed using an ISLA method by removing surviving photon packets able to reach their causal horizons, but this results in an  artificial loss of radiative energy from the source and too low a temperature in the main ionized region. We suggest a compromise solution in Sec.~\ref{subsec:hybrid} below.

\subsection{Far zone}
\label{subsec:farzone}

\subsubsection{Black-body spectra}
\label{subsubsec:fzStars}

The temperature profiles beyond the ionization fronts agree closely between all three codes for both the $T_\mathrm{BB}=10^5$~K and $10^6$~K black-body spectra, including the position of the temperature knee, where the temperature begins its decline to the ambient IGM value. The code results for the temperature begin to depart from each other well beyond the ionized gas region once the temperature declines below $\sim5000$~K. No direct observational consequences are expected.

The ionization structures for hydrogen and helium agree closely well beyond the ionization fronts, with the exception of the direct integration code result for \HeIII, for which the ionization fraction plummets abruptly beyond the \HeIII-front for both black-body spectra. Virtually all of the \HeII-ionizing radiation is absorbed just beyond the \HeIII-front. This appears to be a failing of the scheme. Since \HeIII\ is not directly measured, it has no direct observational consequences.

\subsubsection{Power-law spectra}
\label{subsubsec:fzQSO}

The temperature and ionization structure beyond the ionization fronts agree well between \texttt{PhRay} and \texttt{ENZO} for the $\alpha_Q=0.5$ spectrum for IGM densities at both $z=4$ and $z=6$, although the \texttt{ENZO} temperatures begin to decline somewhat more rapidly at large distances.

For the softer $\alpha_Q=1.73$ spectrum, the \HeII\ fraction from \texttt{ENZO} for the $z=6$ IGM density, while first tracking the \texttt{PhRay} result, suddenly declines at $t\gsim 50$~Myr. The \HeII\ fraction adheres to the \texttt{PhRay} result to greater distances as the spatial resolution is increased for \texttt{ENZO}:\ going from $128^3$ to $256^3$ cells corresponds to decreasing the \HeII\ optical depth at the photoelectric edge from 5.2 to 2.6. As the gas temperature from \texttt{PhRay} remains well above 100~K to distances exceeding 12~Mpc at $t=50$~Myr and 15~Mpc at $t=100$~Myr, \texttt{ENZO} would under-estimate the range around a QSO to which the IGM was heated above the CMB temperature, and so under-estimate the range to which the 21-cm signal would be seen in emission against the CMB around the QSO. In practice, the signal would be complicated by heating from galactic sources, which may well have already warmed the IGM to temperatures above the CMB \citep{2017ApJ...840...39M, 2017MNRAS.471.3632M}.

\subsection{Hybrid RT scheme}
\label{subsec:hybrid}
We develop a hybrid ISLA method applied to sources reionizing their local environment for our revised version of  \texttt{ENZO}, to alleviate the discrepancies in temperature and ionisation structures between the time-dependent RT equation solution and the ISLA solution when removing photon packets that exceed their causal horizon. In the hybrid solution, the propagation speed of the photon packets remains infinite, corresponding to the instantaneous solution of the RT equation, but the travelling distance restriction imposed by causality is enabled only for photons in the sub-luminal region, as given by Eq.~(\ref{eq:sublum}).{\footnote This switch is applied only when the hydrogen around a source is still predominantly neutral, or the helium predominantly neutral or singly ionized. It is also only applied to the photon packets that would effect the reionization. In other situations, the gas will have already been heated by photoionization, with little additional heating from the source, so that no special measures need be taken to ensure an accurate temperature solution. In these cases, the travelling distance restriction is applied to the relevant photon packets to ensure the changing ionization fractions around the source remain causal.}

This hybrid approach captures the accumulated attenuation of the radiation field in the near-luminal expansion region, as the attenuation is mainly at the ionization front. This ensures the gas is heated to approximately the same temperature it would have using time-dependent RT (ie, a finite photon velocity). Once the ionization front slows down to becoming sub-luminal, RT proceeds in a time-independent manner (or quasi-time-dependent allowing for slow changes in the gas or source properties), so that the ISLA method becomes increasingly accurate (Sec. \ref{subsec:QSO}). At the earliest times after the source turns on, however, before the light front reaches the radius at which the ionization front should become sub-luminal, the scheme may produce an ionization front that is acausally large, extending beyond the light front.       
\begin{figure}
    \includegraphics[width=\columnwidth]{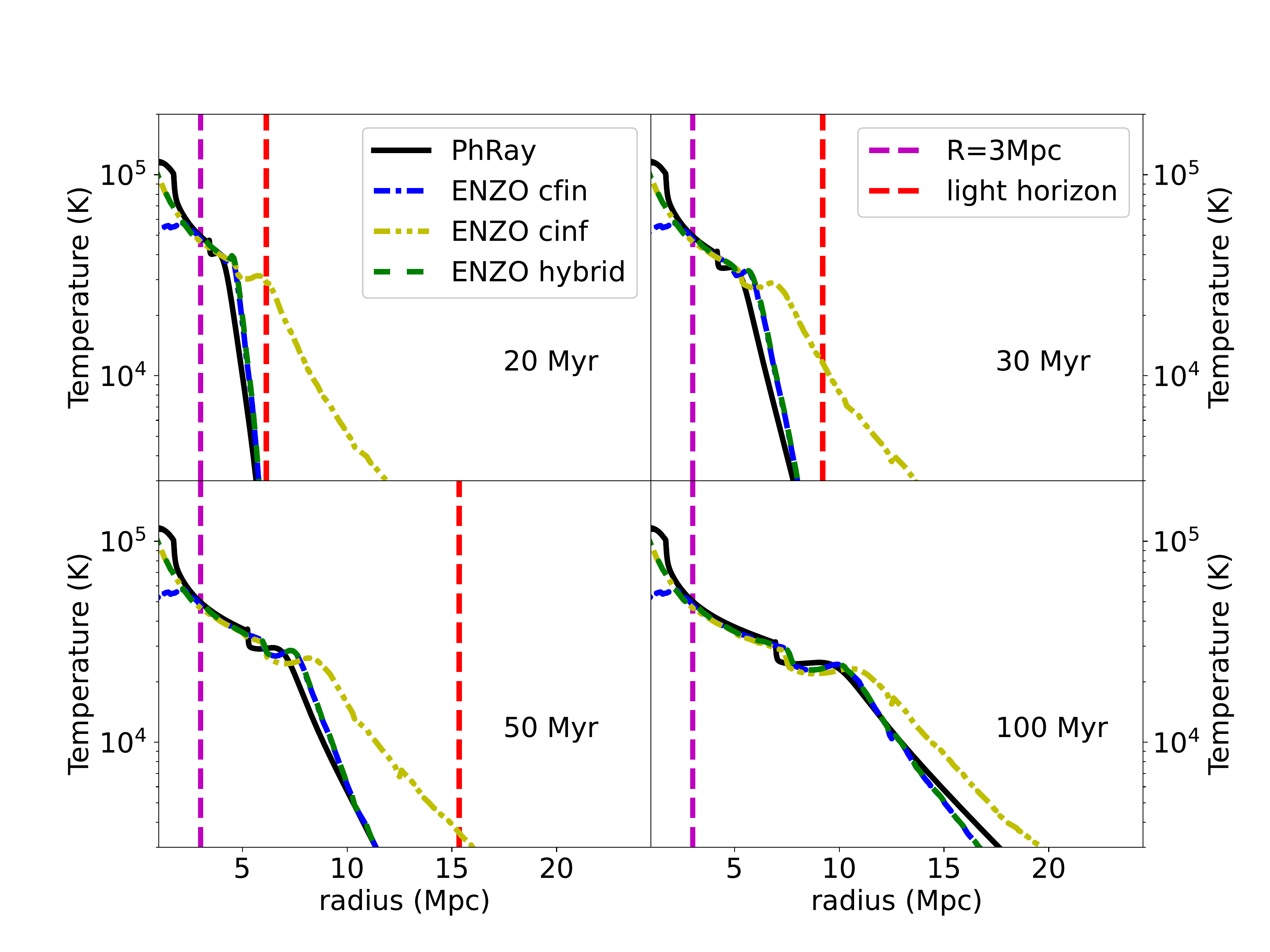}
    \caption{Temperature profiles for reionization at $z=6$ by source $L_{S,\nu}\sim\nu^{-0.5}$. Shown are the results for \texttt{PhRay} (black solid lines), \texttt{ENZO cfin}, for which photon packets are continuously removed when they reach their causal horizon (dot-dashed blue lines), \texttt{ENZO cinf}, for which photon packets are removed only when absorbed or exit the grid (dot-dot-dashed yellow lines) and \texttt{ENZO hybrid} (dashed green lines), for which photons are removed when they reach their causal horizon only if they are located in the sub-luminal region ($R>3\,\rm Mpc$, shown by the vertical dashed magenta line). The dashed red line in each panel shows the position of the light horizon.}
    \label{fig:powerlaw_0.5_z6_Temperature_hybrid}
\end{figure}

\begin{figure}
    \includegraphics[width=\columnwidth]{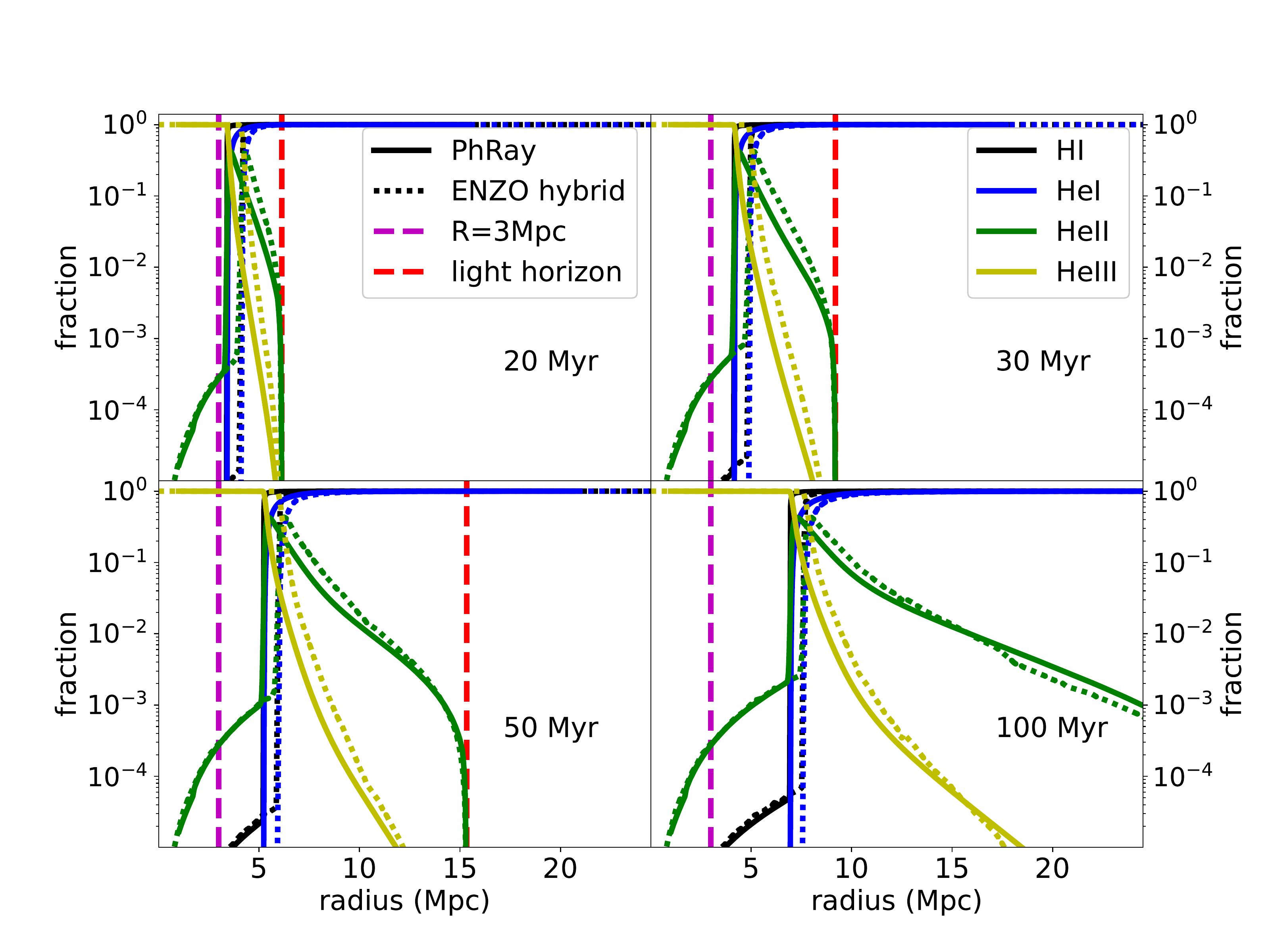}
    \caption{Ionisation profiles for reionization at $z=6$ by source $L_{S,\nu}\sim\nu^{-0.5}$. Shown are the results for \texttt{PhRay} (solid lines) and \texttt{ENZO hybrid} (dotted lines), for which photon packets reaching their causal horizon are removed only in the sub-luminal region ($R>3\,\rm Mpc$, shown by the vertical dashed magenta line), for the \HI\ (black lines), \HeI\ (blue lines), \HeII\ (green lines) and \HeIII\ (yellow lines) ionization fractions. The dashed red line in each panel shows the position of the light horizon.}
    \label{fig:powerlaw_0.5_z6_ion_hybrid}
\end{figure}

Fig.\ref{fig:powerlaw_0.5_z6_Temperature_hybrid} and Fig.\ref{fig:powerlaw_0.5_z6_ion_hybrid} illustrate the temperature and ionisation profiles for reionization at $z = 6$ by a source $L_{S,\nu}\sim\nu^{-0.5}$ computed using three methods:\ (1)\ the ISLA method (case \lq cinf\rq), (2)\ removing photons everywhere when they exceed their light horizon (case \lq cfin\rq), and (3)\ the hybrid scheme. The ionization zone is too large in the ISLA method. Removing photons everywhere when they exceed their causal radius results in too great an energy loss in the near luminal expansion region, with the resulting temperature too low in the region. For the hybrid method (\texttt{ENZO} hybrid), the agreement of the temperature and ionisation structures with those of the time-dependent RT solution from \texttt{PhRay} is much improved not only in the far zone but in the near zone as well. For the softer $\alpha_Q=1.73$ spectrum, we find improved agreement by defining the sub-luminal region according to the radius at which the expansion speed of the ionization front declines to $0.2c$ instead of $0.5c$. Interpolation may be used for intermediate values of $\alpha_Q$. These choices may be applied for each source individually in a multiple-source simulation with a range of source spectra, although fixing the radius according to an ionization front speed of $0.5c$ may be adequate, as the temperature differences between the static and time-dependent RT solutions are smaller for softer spectra.

\subsection{Cosmological simulation application}
\label{subsec:cosmo_sim}
We apply the three different ISLA methods (\texttt{ENZO cinf}: the original ISLA, \texttt{ENZO cfin}: ISLA, but adopting the causal travel distance restriction throughout the entire simulation volume and \texttt{ENZO hybrid}: ISLA, but applying the causal travel distance restriction only in the sub-luminal ionization front expansion region) to a cosmological hydrodynamic simulation using our revised version of \texttt{ENZO} to study the temperature and ionization structure around a QSO. Assuming for simplicity that no metagalactic ultraviolet background (UVB) is present, we turn on a beamed QSO-like radiation source with an $\alpha_Q = 0.5$ power-law spectrum at the centre of the simulation box at $z = 7$. The total hydrogen-ionizing photon emission rate is $\dot N_\gamma = 1.5\times10^{57}\,\mathrm{s}^{-1}$ and the opening angle of the source is $10^{\circ}$. The cosmological parameters assumed are $\Omega_{m} = 0.27$,  $\Omega_{b} = 0.046$, $\Omega_{\Lambda} = 0.73$, $h = 0.70$, $\sigma_{8} = 0.811$ and $n_{s} = 0.961$, with a primordial helium mass fraction $Y = 0.24$, consistent with \emph{PLANCK} measurements \citep{2018arXiv180706209P}. The code is run in unigrid mode with a comoving box size of $120 h^{-1}\,\rm Mpc$ and $256^{3}$ cubic cells. (The spatial resolution in proper units at $z = 7$ is comparable to the spatial resolution in the test problems in Sec.\ref{subsec:QSO}.) The Cold Dark Matter initial conditions at $z=50$ are generated by the \textbf{MUSIC} code \citep{2011MNRAS.415.2101H}; the code also sets the baryon properties, with a low temperature given by adiabatic expansion following the recombination epoch. The chemical and cooling processes are computed by \textbf{GRACKLE} \footnote{https://grackle.readthedocs.io/} \citep{2017MNRAS.466.2217S}. The RT equations are solved in a sub-cycle process in \texttt{ENZO}, so that the cosmological simulations are fully coupled radiation hydrodynamics simulations, rather than being performed as a post-processing step, like in most cosmological RT simulations \citep[eg][]{2002MNRAS.332..601S, 2004MNRAS.348L..43B, 2009ApJ...694..842M, 2012MNRAS.423..558C, 2013MNRAS.435.3169C, 2017MNRAS.468.3718K, 2018MNRAS.476.1174E, 2020MNRAS.498.6083E}.

\begin{figure}
    \includegraphics[width=\columnwidth]{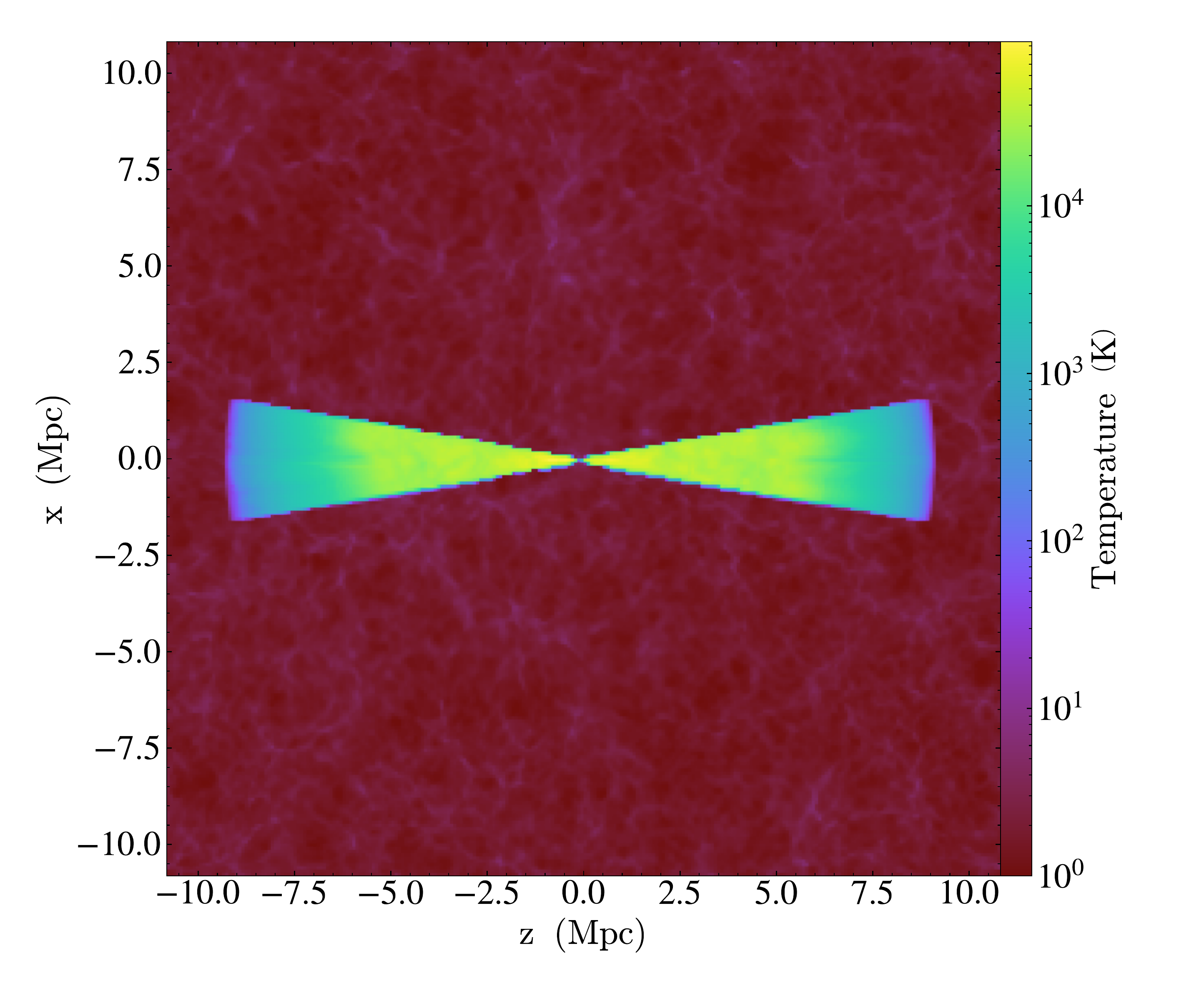}
    \caption{Temperature slice plot at $z \sim 6.93$ for a beamed QSO-like radiation source ($L_{S,\nu}\sim\nu^{-0.5}$) after $30\, \rm Myr$. The \texttt{ENZO hybrid} RT method is used, for which the causal travel distance restriction for photon packets is applied only in the sub-luminal ionization front expansion region (see text).
    }
    \label{fig:temperature_slice_30Myr}
\end{figure}

\begin{figure}
    \includegraphics[width=\columnwidth]{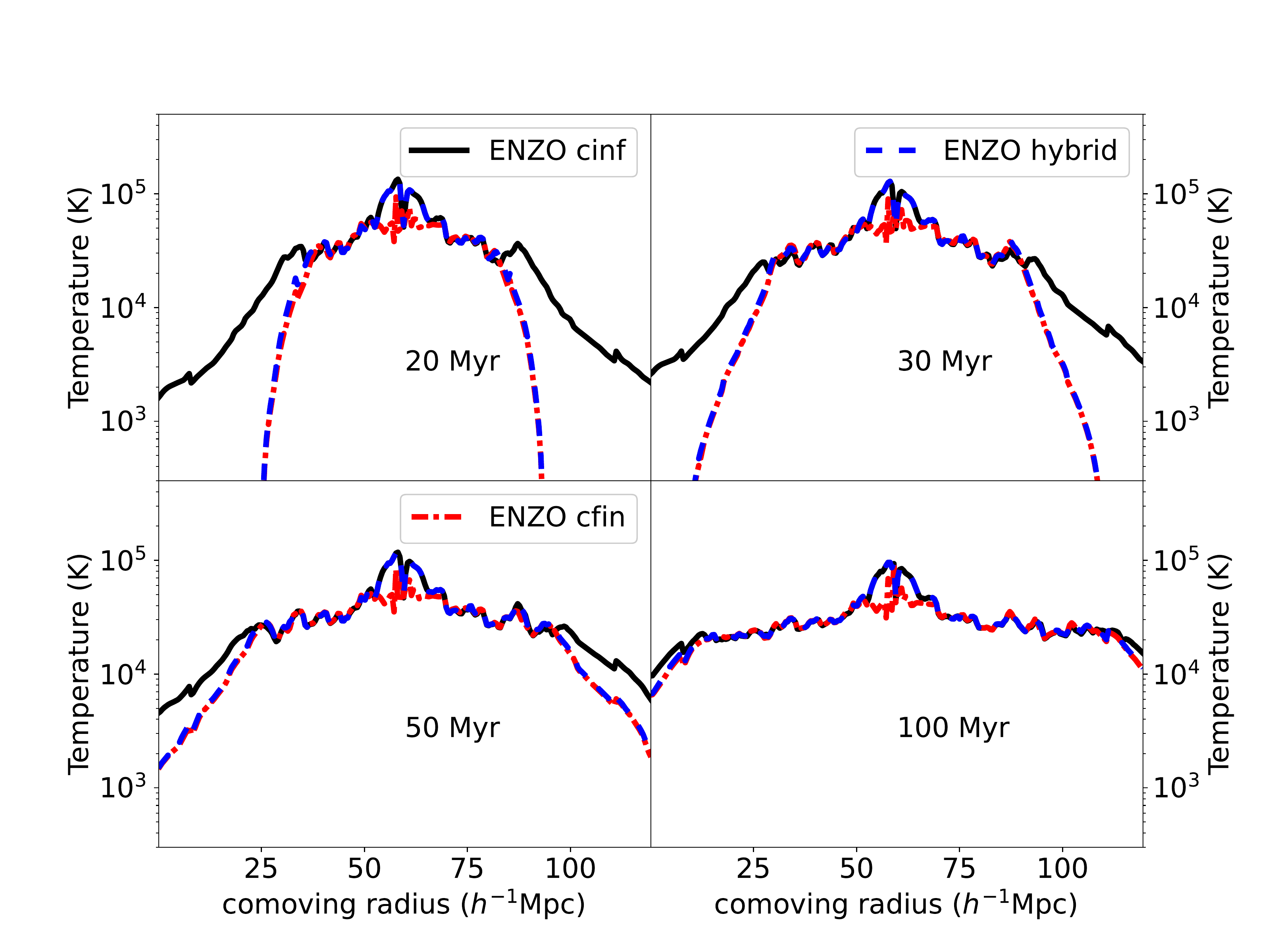}
    \caption{Temperature profiles for a QSO-like radiation source ($L_{S,\nu}\sim\nu^{-0.5}$) at the indicated times after the source turns on at $z=7$:\ $20\, \rm Myr$, $30\, \rm Myr$, $50\, \rm Myr$ and $100\, \rm Myr$. The radiation source is located at the centre of the line of sight. Results are shown for the \texttt{ENZO cinf} method (solid black lines), the \texttt{ENZO cfin} method (dot-dashed red lines) and for \texttt{ENZO hybrid} (dashed blue lines). Temperature profiles in the central luminal ionization front expansion region are underestimated by \texttt{ENZO cfin}. The temperature discrepancy in the region surrounding the radiation source is as high as $5 \times 10^4$~K. }
    \label{fig:temperature_profile_cosmic}
\end{figure}

\begin{figure}
    \includegraphics[width=\columnwidth]{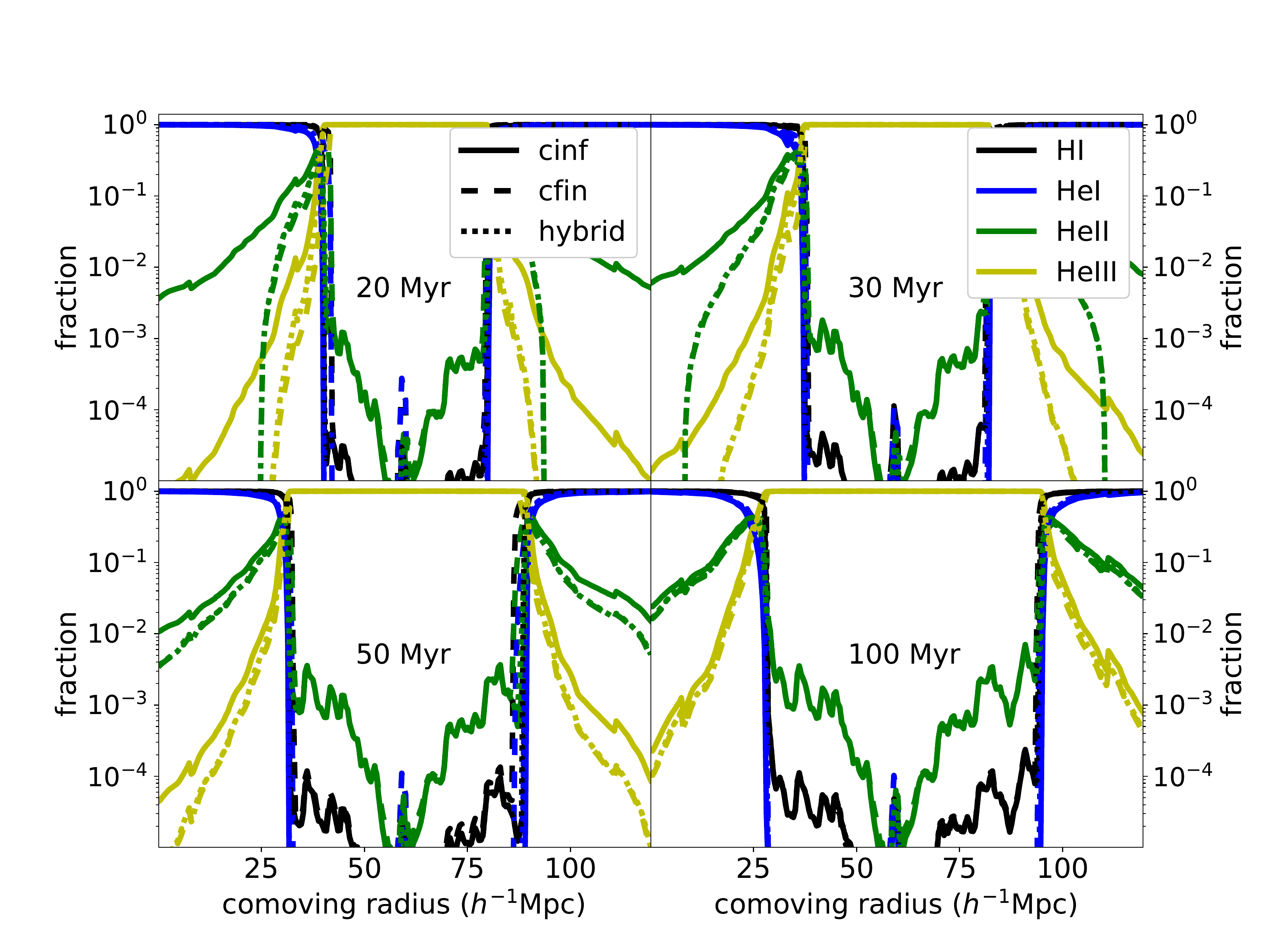}
    \caption{Ionisation profiles for the QSO-like radiation source ($L_{S,\nu}\sim\nu^{-0.5}$) at times $20\, \rm Myr$, $30\, \rm Myr$, $50\, \rm Myr$ and $100\, \rm Myr$. The radiation source is located at the centre of the line of sight. Results are for the \texttt{ENZO cinf} method (solid lines), \texttt{ENZO cfin} method (dashed lines) and for \texttt{ENZO hybrid} (dotted lines).}
    \label{fig:ionised_profile_cosmic}
\end{figure}

\begin{figure}
    \includegraphics[width=\columnwidth]{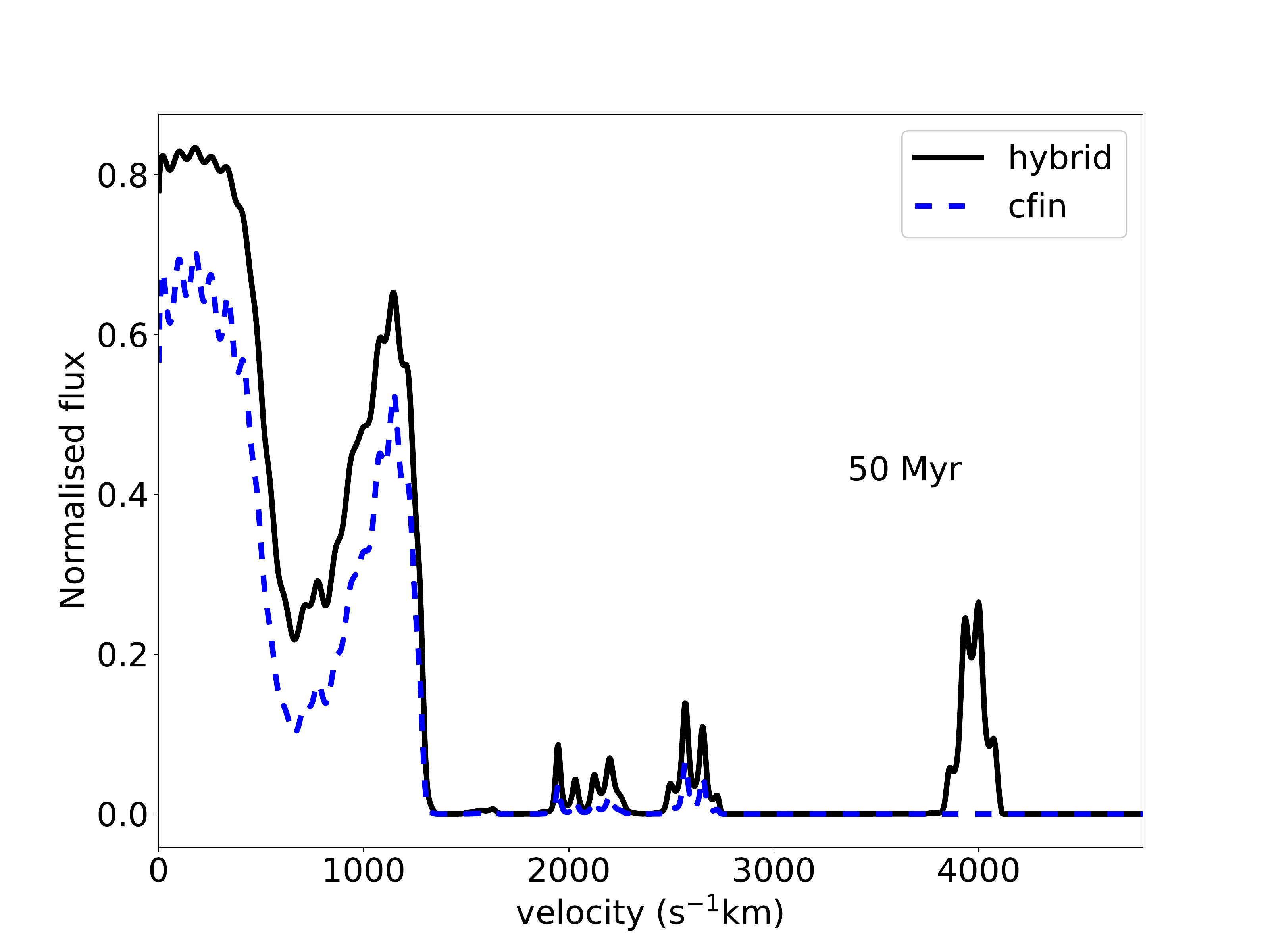}
    \caption{Normalised $\rm Ly_{\rm \alpha}$ flux spectra emitted by a QSO-like radiation source 50~Myr after the source turns on at $z=7$, allowing for intervening absorption along the line of sight. Results are shown for \texttt{ENZO hybrid}  (solid black line) and for \texttt{ENZO cfin} (dashed blue line). (The \Lya\ absorption spectra are computed following \citet{2019MNRAS.484.4273L}.)}
    \label{fig:LY_alpha_profile}
\end{figure}

The temperature around the source is shown in  Fig.~\ref{fig:temperature_slice_30Myr}, with the temperature declining with distance from the source, and modulated by the large-scale structure of the gas. The sharp ends to the temperature cone correspond to the light fronts. In Fig.~\ref{fig:temperature_profile_cosmic}, the temperature along a line of sight through the beam centre is shown for the three methods. The ISLA method (\texttt{ENZO cinf}) produces high excess temperatures away from the source. The \texttt{ENZO cfin} and \texttt{ENZO hybrid} methods agree in temperature on large scales, but the temperature from the \texttt{ENZO cfin} method is too low in the luminal ionization expansion region, within the inner 2~Mpc from the source, by up to $5\times10^4$~K. The different predictions for the ionization fractions are shown in Fig.~\ref{fig:ionised_profile_cosmic}. The ISLA method again gives excess ionization on large scales. The \texttt{ENZO cfin} and \texttt{ENZO hybrid} methods agree, except within the inner 2~Mpc. The near zone \Lya\ forest is shown in Fig.~\ref{fig:LY_alpha_profile}. The lower temperatures using \texttt{ENZO cfin} result in a larger radiative recombination rate and so a greater amount of absorption near the source. Whilst the result from \texttt{ENZO hybrid} more faithfully recovers the expected gas temperature within the luminal region (Sec.~\ref{subsec:hybrid}), the actual amount of absorption would be still somewhat smaller in this region because of the temperature boost allowing for the time-dependent RT in the luminal zone. For precision work, a time-dependent solution to the RT equation would be required for such a hard spectrum. The discrepancy is smaller for a softer spectrum (Sec.~\ref{subsubsec:reion}).

\section{Conclusions}
\label{sec:conclusions}

We compare three radiative transfer codes applied to photoionization problems for sources with spectra typical of stars (black body) and QSOs (power law). One code integrates the time-independent radiative transfer equation directly, and is applied only to the black-body spectra problems. The other two use photon packets to solve for the radiative transfer, one assuming instantaneous photoionization (with the distance photon packets travel limited by the speed of light) and the other retaining fully the time-dependent term in the radiative transfer equation. Our main findings are:

1.\ Photon packet codes are far more efficient at solving the radiative transfer problem for photoionization compared with direct integration. Fewer photon frequencies and coarser spatial gridding are tolerated by the photon packet codes, with optical depths at the threshold energy able to exceed unity with accurate solutions. Another shortcoming of the direct integration code is that it may fail to propagate low levels of doubly ionized helium beyond the \HeIII-front as far as do the photon packet codes.

2.\ All methods agree well on the growth of the nearly fully ionized regions, although the ionization fronts from the direct integration scheme tend slightly to lead those from the photon packet codes. The successful solution of the ionized regions is a significant achievement of the photon packet codes particularly for hydrogen ionization, as the spatial grid used for the instantaneous photoionization version corresponds to a hydrogen optical depth per grid zone exceeding 100 at the photoelectric threshold. We recommend, however, that for ionizing singly ionized helium, the optical depth at the singly ionized helium threshold should be close to unity or smaller. 

3.\ Including the time-dependent differential operator in the radiative transfer equation is essential when ionization fronts expand near the speed of light. Solutions to the radiative transfer equation in the infinite-speed-of-light approximation (corresponding to solving the time-independent RT equation) may substantially under-estimate the temperature in these regions. The under-estimate increases with the hardness of the spectrum, with the temperature discrepancy exceeding $5\times10^4$~K for gas that was initially neutral, as may arise for reionization at high redshifts by QSOs with hard spectra. A scheme that solves the time-dependent RT equation is thus required to obtain an accurate solution in the near zones of QSOs that photoionize the IGM.

4.\ The boost in temperature due to time-dependent RT is larger when both hydrogen and helium are initially predominantly neutral compared with the case when the hydrogen is predominantly ionized and the helium singly ionized, as may arise when the gas is initially ionized by a metagalactic UV background radiation field dominated by galactic sources.

5.\ Outside the luminal expansion region, the gas temperature and ionization structure agree well between the time-dependent and infinite-speed-of-light photon packet codes, although some differences arise at large distances where the gas is predominantly neutral. These differences appear to result from differences in spatial resolution, rather than from the assumption of an infinite speed of light.

6.\ A photon packet code recovers the correct solutions to the time-dependent RT equation for an ionization front to good approximation using a hybrid scheme. In this scheme, the RT equation is solved in the infinite-speed-of-light approximation only out to the radius at which the velocity of the ionization front declines to approximately half the speed of light. Photon packets outside this radius are removed if they travel to distances beyond the light front of the source.

7.\ Photon energies well above the photoionization thresholds must be included to capture the warming of the largely neutral gas well outside the ionization regions for power-law spectra. The required maximum photon energy increases for softer spectra.

\section*{Acknowledgments}
The authors thank B. Smith and J. Wise for helpful conversations, and the referee for numerous suggestions to improve the manuscript, including the suggestion to make direct comparisons between our results and the published literature. KHL acknowledges financial support from the School of Physics and Astronomy, University of Edinburgh. KHL thanks the Computational Astrophysics Lab at National Taiwan University for support. KHT thanks the Robert Cormack Bequest fund for a Summer Vacation Research Scholarship. Computations described in this work were performed using the \texttt{ENZO} code developed by the Laboratory for Computational Astrophysics at the University of California in San Diego (http://lca.ucsd.edu).

\section*{Data Availability}
No new observational data were generated or analysed in support of this research.


\bibliographystyle{mnras}
\bibliography{mnras_template}

\appendix

\section{Comparisons with test problems in the literature}
\label{appendix:comparisons}

\begin{table}
\begin{center}
\begin{tabular}{|l|c|c|c|c|c|}\hline
Ref &$\dot N_\gamma$ & $\alpha_Q$ & $h\nu_\mathrm{max}$ & $n_\mathrm{H}$  & $z$\\
& s$^{-1}$ & & keV & cm$^{-3}$  \\
\hline
\hline
AH99 & $3\times10^{56}$ & 1.8 & $\infty$ & $5.9\times10^{-5}$   &6 \\ \hline
D16 & $10^{57}$ & 1.5 & 2.17 & $9.7\times10^{-5}$  &7  \\ \hline
G18 & $1.36\times10^{56}$ & 1.5 & 3 & $9.6\times10^{-5}$  &7\\ \hline
CG21 & $10^{57}$ & 1.5 & 1 & $9.7\times10^{-5}$ &7  \\ \hline
\end{tabular}
\end{center}
\caption{Parameters for the test problems of \citet{1999ApJ...520L..13A} (AH99), \citet{2016MNRAS.457.3006D} (D16), \citet{2018MNRAS.479.4320G} (G18) and \citet{2021ApJ...911...60C} (CG21).
}
\label{tab:test_prob_params}
\end{table}

Solutions of the time-dependent radiative transfer equation for power-law spectra using \texttt{PhRay} are compared with published test problems of the photoionization of hydrogen and helium using ISLA methods, as provided by \citet{1999ApJ...520L..13A}, \citet{2016MNRAS.457.3006D}, \citet{2018MNRAS.479.4320G} and \citet{2021ApJ...911...60C}. The thermodynamics are governed by photoelectric heating and radiative cooling, including inverse Compton cooling off the Cosmic Microwave Background at the indicated redshift. Secondary electron ionizations and associated energy losses are included as indicated, using the fits from \citet{2002ApJ...575...33R} to the Monte Carlo computations of \citet{1985ApJ...298..268S}. Parameters for the test problems are provided in Table~\ref{tab:test_prob_params}, showing the net hydrogen ionizing photon production rate $\dot N_\gamma$, the power-law exponent $\alpha_Q$ for the QSO spectrum, the upper photon energy cutoff $h\nu_\mathrm{max}$ for the spectrum, the hydrogen density $n_\mathrm{H}$ of the surrounding gas and the redshift, which controls the inverse Compton cooling rate.

\begin{figure}
    \centering
    \includegraphics[width=\columnwidth]{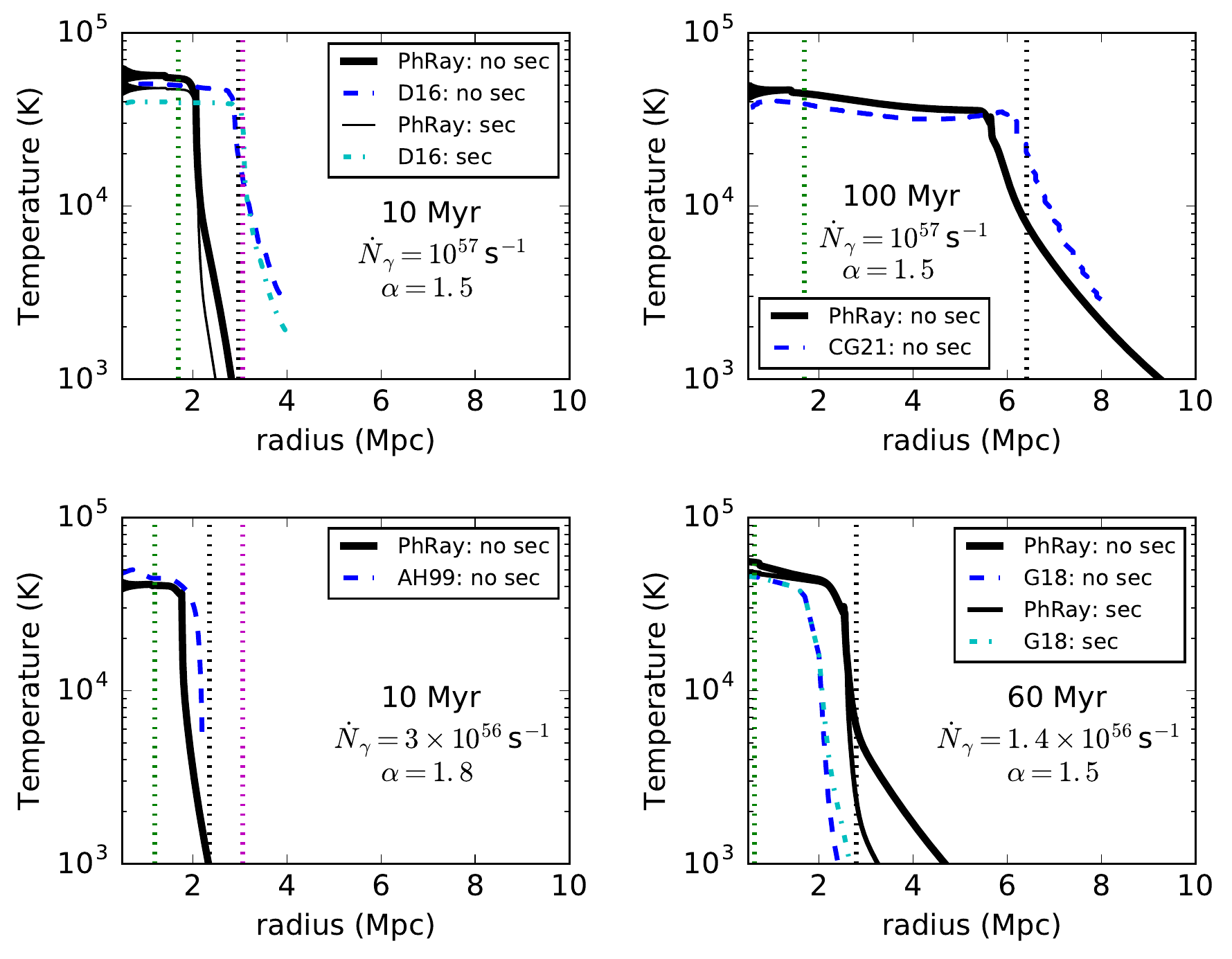}
    \caption{Comparison between solutions to test problems provided by \texttt{PhRay} using a finite speed of light and published results using ISLA methods. The vertical green dashed lines indicate the theoretical position of the hydrogen ionisation front when its advance slows to speed $c/2$. The vertical black dotted lines show the maximum possible radius of the hydrogen ionization front. The vertical magenta dotted lines show the position of the light front. In the top left and bottom right panels, solutions both without (\lq no sec\rq) and with (\lq sec\rq) secondary electron ionizations are shown; for the other panels, the published results did not include secondary electron ionizations.}
    \label{fig:QSO_ReI_lit_Temp_compare}
\end{figure}

The top left hand panel of Fig.~\ref{fig:QSO_ReI_lit_Temp_compare} compares the solutions from \texttt{PhRay} without and with secondary ionizations to those provided by \citet{2016MNRAS.457.3006D} using a spherically symmetric 1D ISLA method for a QSO spectrum with $\alpha_Q=1.5$. The medium surrounding the source is assumed static in this problem. The ISLA solutions from \citet{2016MNRAS.457.3006D} (blue dashed and cyan dot-dashed lines) extend some distance beyond the solution of the time-dependent RT equation using the correct speed of light (thick and thin black solid lines) given by \texttt{PhRay}, and even beyond the light front (shown as the vertical magenta dotted line). The central, main ionized region from \citet{2016MNRAS.457.3006D} reaches the maximum possible radius of the \HII-front, given by Eq.~(\ref{eq:rhii})\footnote{Here and for the other test problems, the full value for $\dot N_\gamma$ is used. Since helium also absorbs photons above the helium ionization thresholds, the maximum radius will be somewhat smaller.} (vertical black dotted line), which is nearly coincident with the light front (vertical magenta dotted line). Keeping up with the maximum radius is expected for an ISLA scheme, which allows photons to travel until absorbed, but for the ionization front to have reached this distance, it had to travel superluminally at earlier times. By contrast, at $10$~Myr the dominant ionized region from \texttt{PhRay}, where $T>10^4$~K, extends just beyond the distance to which the \HII-front travels at a speed exceeding $c/2$, as given by Eq.~(\ref{eq:sublum}) (shown by the vertical green dotted line). In the central, main ionized region, the temperatures agree well between the two calculations, although the temperatures from \texttt{PhRay} are somewhat higher by about 10\%. Significant cooling is provided by secondary electron ionization losses both in the central ionized region and in the extended region where the gas temperature is below $10^4$~K.

The top right hand panel compares the solutions from \texttt{PhRay} and  \citet{2021ApJ...911...60C}, who use an algorithm very similar to that of \citet{2016MNRAS.457.3006D} for solving the static 1D RT problem, although modified to allow for a variable timestep within regions of very different ionization levels. No secondary electron ionizations are allowed for, and the surrounding medium is assumed static in this problem. The ISLA solution of \citet{2021ApJ...911...60C} has an \HII-front that extends nearly to its maximum possible radius (vertical black dotted line), and is somewhat beyond that obtained by \texttt{PhRay}. The difference reflects an earlier superluminal expansion phase of the \HII\ region in the ISLA computation. In the inner main ionized region, the temperature from \texttt{PhRay} mildly exceeds that obtained by \citet{2021ApJ...911...60C} by about 15\%.

In the lower left panel, the result for a steeper spectrum ($\alpha_Q=1.8$) from \texttt{PhRay} is compared with the ISLA solution of \citet{1999ApJ...520L..13A}. Adiabatic cooling is included in this problem, although it negligibly affects the temperature over the brief interval of 10~Myr of the computation. Secondary electron ionizations are not accounted for in the problem. The temperatures agree well in the central region, with a slightly higher temperature obtained by \citet{1999ApJ...520L..13A}. The ISLA solution also has a slightly advanced \HII-front compared with the time-dependent RT solution from \texttt{PhRay}, more nearly reaching its maximum possible radius (vertical black dotted line). The reason the central temperature from the ISLA solution slightly exceeds that of the time-dependent RT solution is unclear; the grid and frequency resolution are not provided by \citet{1999ApJ...520L..13A} and there is no description of convergence tests on either.

Lastly, in the lower right panel we compare the solution of \texttt{PhRay} with the ISLA solution of \citet{2018MNRAS.479.4320G}. Computations both without and with secondary electron ionizations were performed. The solutions of \citet{2018MNRAS.479.4320G} are anomalous in that the ionization fronts advance too slowly. From Eq.~(\ref{eq:rhii}), the \HII-front should be located at $r_\mathrm{HII}\simeq2.8$~Mpc (vertical black dotted line), in good agreement with the result from \texttt{PhRay}, whilst \citet{2018MNRAS.479.4320G} find the front to be located at $\sim2.0$~Mpc. Even allowing for all photons above the \HeI\ threshold to be absorbed by helium atoms, a production rate of purely hydrogen-ionizing photons of $8.0\times10^{55}$~s$^{-1}$ would remain, giving an \HII-front position of 2.3~Mpc. As the radiative recombination time is longer than $10^9$~yr, this position should have been reached. Another anomaly is their temperature allowing for secondary electron ionization energy losses, which unexpectedly exceeds the temperature without secondary electron ionization losses, contrary to the result from \texttt{PhRay}. The discrepancies may be consequences of the large optical depths per grid zone in the computation of \citet{2018MNRAS.479.4320G}. At the photoelectric edges, the optical depths before the gas is photoionized are $\sim70$ for \HI\ and $\sim1.4$ for \HeII. The high \HI\ optical depth may result in too little penetration of ionizing photon packets into the still neutral gas. By comparison, for another simulation in \citet{2018MNRAS.479.4320G} of a QSO embedded in a halo with higher spatial resolution, the \HI\ and \HeII\ optical depths at the average IGM gas density are $\sim16$ and 0.3, respectively, and the size of the \HII\ region found is in good agreement with the analytic estimate.

We also ran \texttt{PhRay} on a test problem with a $10^5$~K black body spectrum source emitting at a hydrogen-ionizing photon rate $10^{51}$~s$^{-1}$ into a static medium with a cosmic abundance of hydrogen and helium, hydrogen density $n_\mathrm{H}=0.1$~cm$^{-3}$ and initial temperature 100~K, to compare with Test 1, without metals but with secondary electron ionizations, of \citet{2013MNRAS.431..722G}. After $10^5$~yr, the temperature $50$~pc from the source is $4.3\times10^4$~K, declining gradually to $4.0\times10^4$~K at 100~pc, $3.2\times10^4$~K at 200~pc and $2.3\times10^4$~K at 500~pc. The temperatures are comparable to, but slightly in excess by about 0.1~dex of, the temperatures from the \texttt{CLOUDY} ionization code reported by \citet{2013MNRAS.431..722G}. From Eq.~(\ref{eq:sublum}), the ionization front for this problem will expand at a speed exceeding $c/2$ until it reaches 5~kpc, so the slightly higher temperatures are expected since \texttt{CLOUDY} is not designed to track the relaxation of temperatures following the heating by near luminal \HII-front expansion to their steady-state value.

\section{Convergence tests}
\label{appendix:convergence}


We show convergence tests for QSO reionization simulations which are performed by \texttt{ENZO} v2.6. For all the convergence tests, we adopt identical parameters relating to the ray-tracing method of \texttt{ENZO}. In particular, the minimum ray angular resolution parameter is $\Phi_c = 15.1$ and the HEALPix Level is 6
\citep[see the definitions in][]{2011MNRAS.414.3458W}. The spacial resolution is the only code parameter varied for the convergence tests. We also use an identical set of energy intervals and energy bins for simulations with various power-law indices. For all the power-law spectra, the energy interval ranges from $13.6- 1000.0\,\mathrm{eV}$ and the selected energy bins are $[14.12, 16.13, 19.10, 22.06, 24.08, 26.00, 31.48, 39.50, 47.52,\\ 53.00,66.74, 118.20, 205.98, 322.29, 456.81, 597.59, 732.11,\\ 848.42, 936.20, 987.66]\,\mathrm{eV}$.

\begin{figure}
    \centering
    \includegraphics[width=\columnwidth]{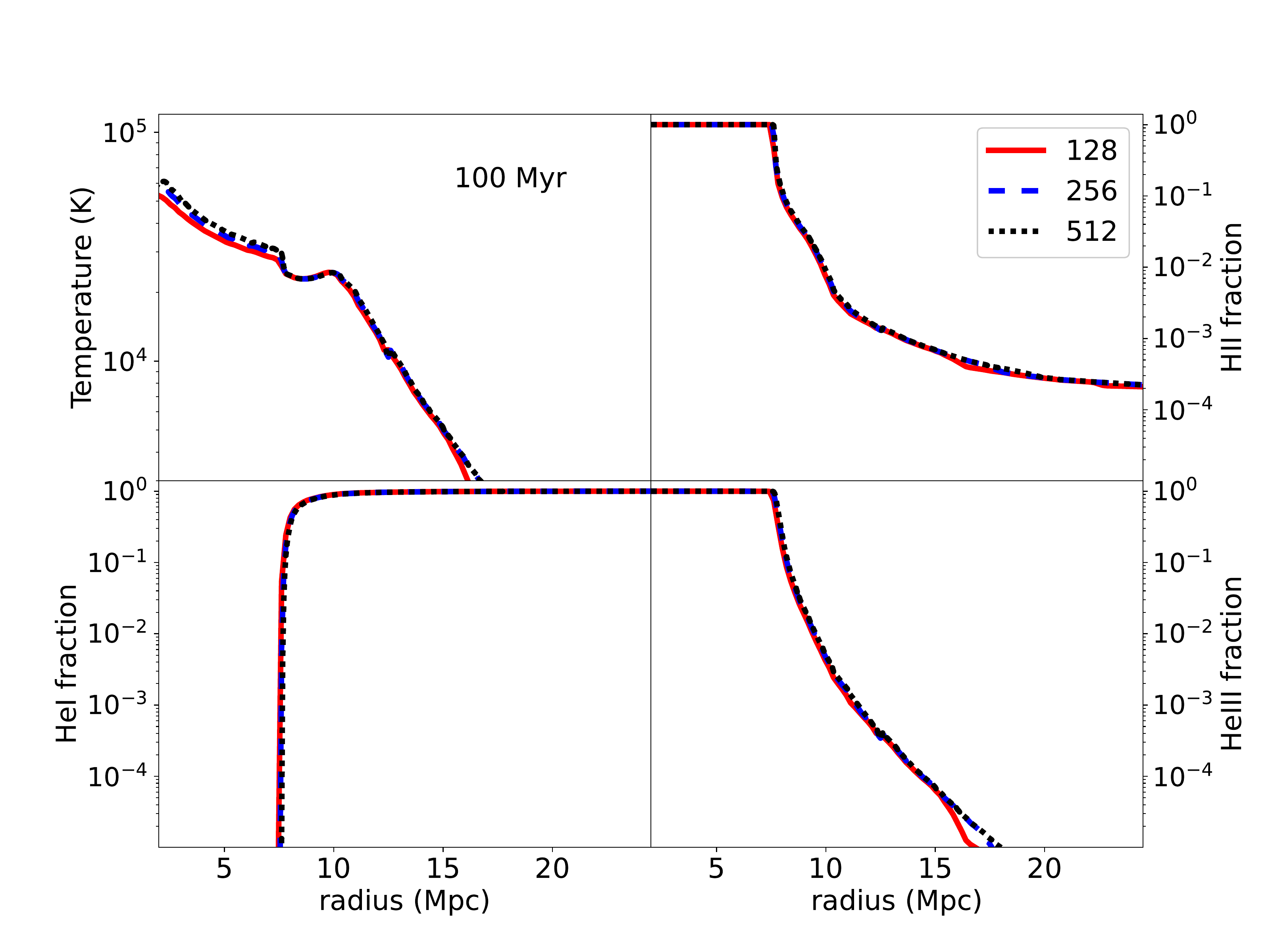}
    \caption{Profiles at $100$ Myr for reionization at $z=6$ by a source $L_{S,\nu}\sim\nu^{-0.5}$. The upper left panel shows that the relative difference in temperature between $(25$ Mpc, $256^{3})$ (blue dashed line) and $(25$ Mpc, $512^{3})$ (black dotted line) simulations is within $10\%$ in the fully ionized region.}
    \label{fig:QSO_0.5_6}
\end{figure}

\begin{figure}
    \includegraphics[width=\columnwidth]{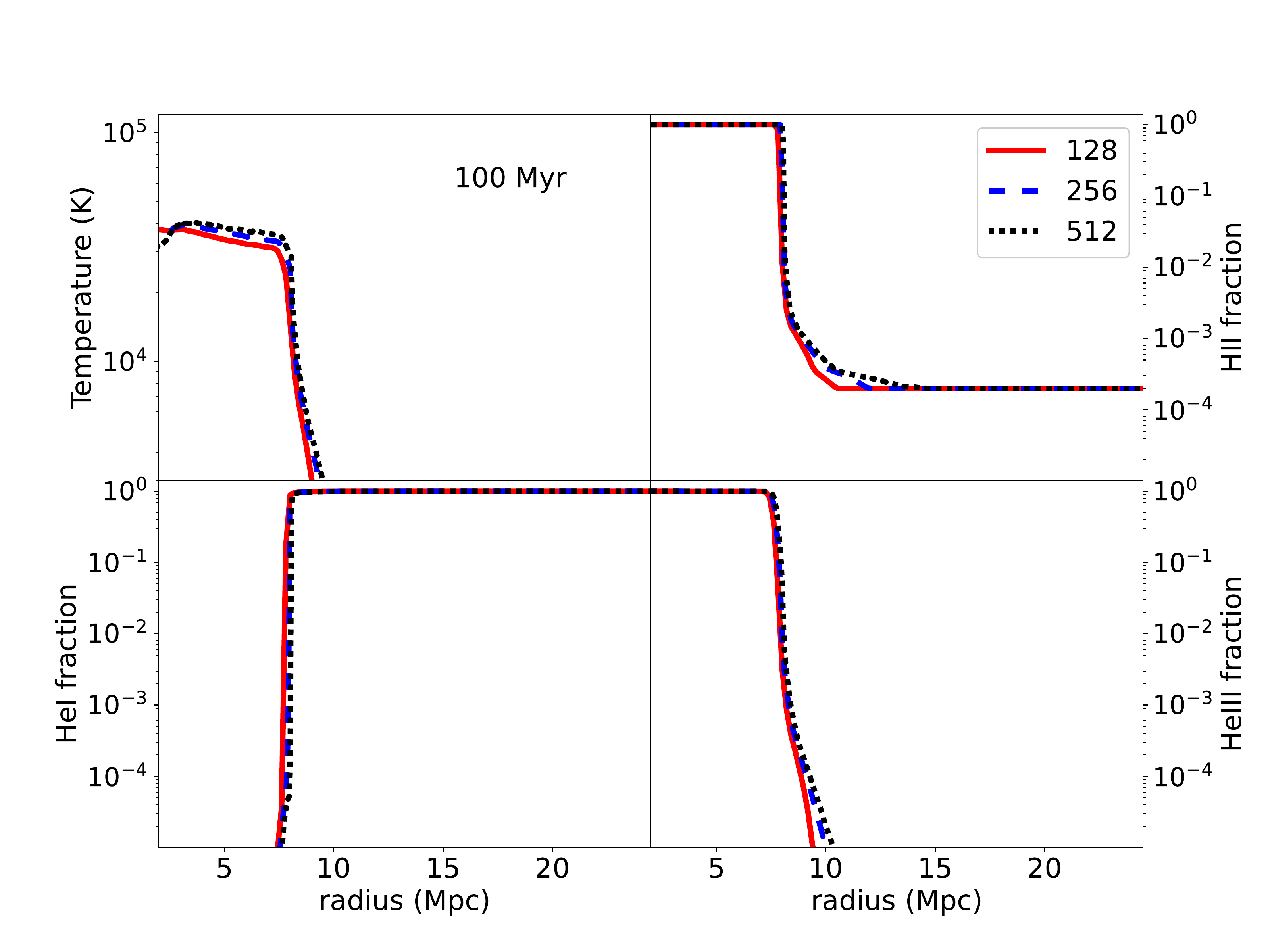}
    \caption{Profiles at $100$ Myr for reionization at $z=6$ by a source $L_{S,\nu}\sim\nu^{-1.73}$. The upper left panel shows that the relative difference of temperature between $(25$ Mpc, $256^{3})$ and $(25$ Mpc, $512^{3})$ simulations is within $10\%$ in the fully ionized region. All the panels show that increasing the resolution of the simulations slightly advances the positions of the ionisation fronts.}
    \label{fig:QSO_1.7_6}
\end{figure}

\begin{figure}
    \includegraphics[width=\columnwidth]{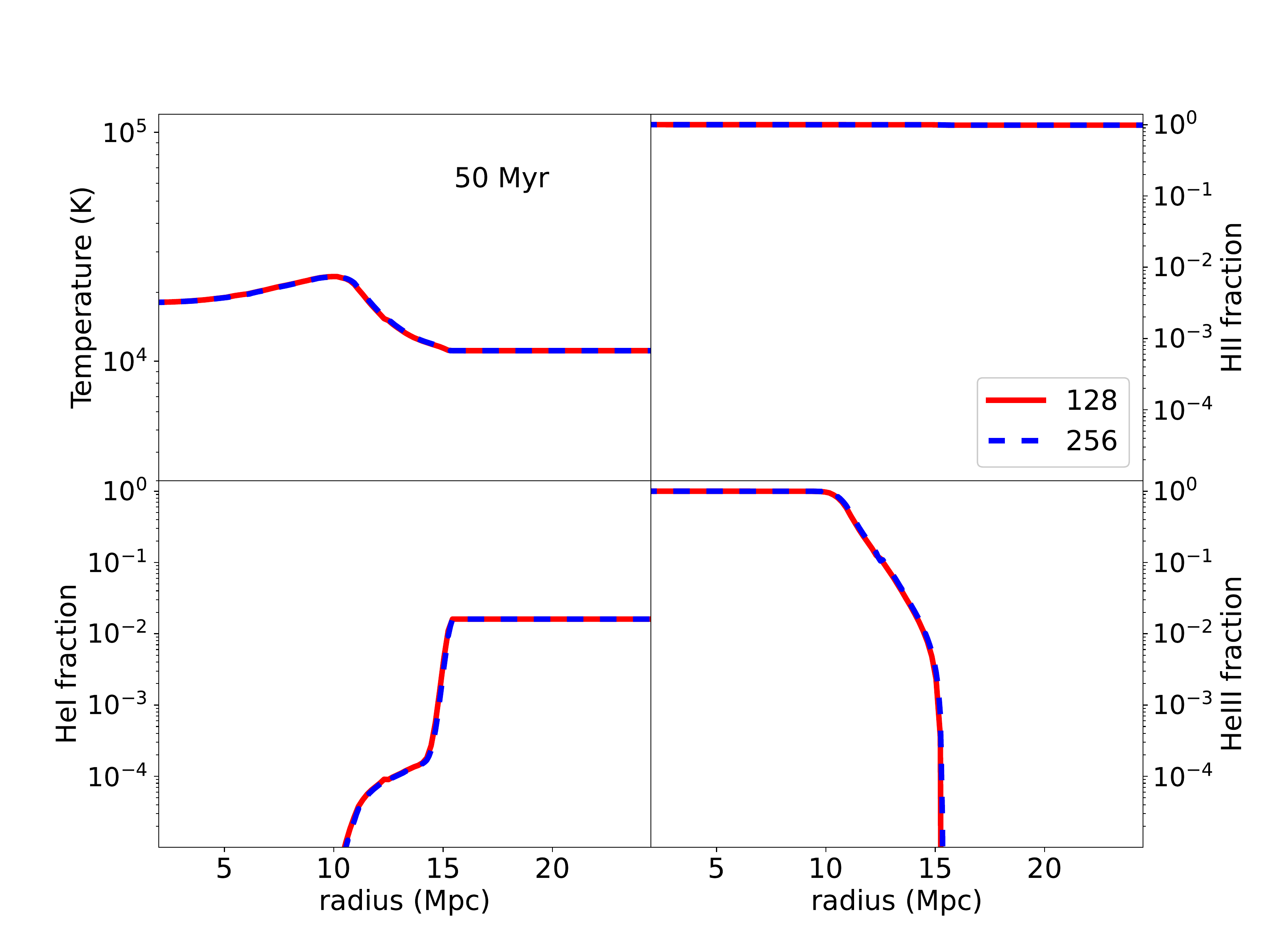}
    \caption{Profiles at 50 Myr for reionization at $z=4$ by a source $L_{S,\nu}\sim\nu^{-0.5}$. All the quantities are converged in these $(25$ Mpc, $128^{3})$ and $(25$ Mpc, $256^{3})$ simulations.}
    \label{fig:QSO_0.5_4}
\end{figure}

\begin{figure}
    \includegraphics[width=\columnwidth]{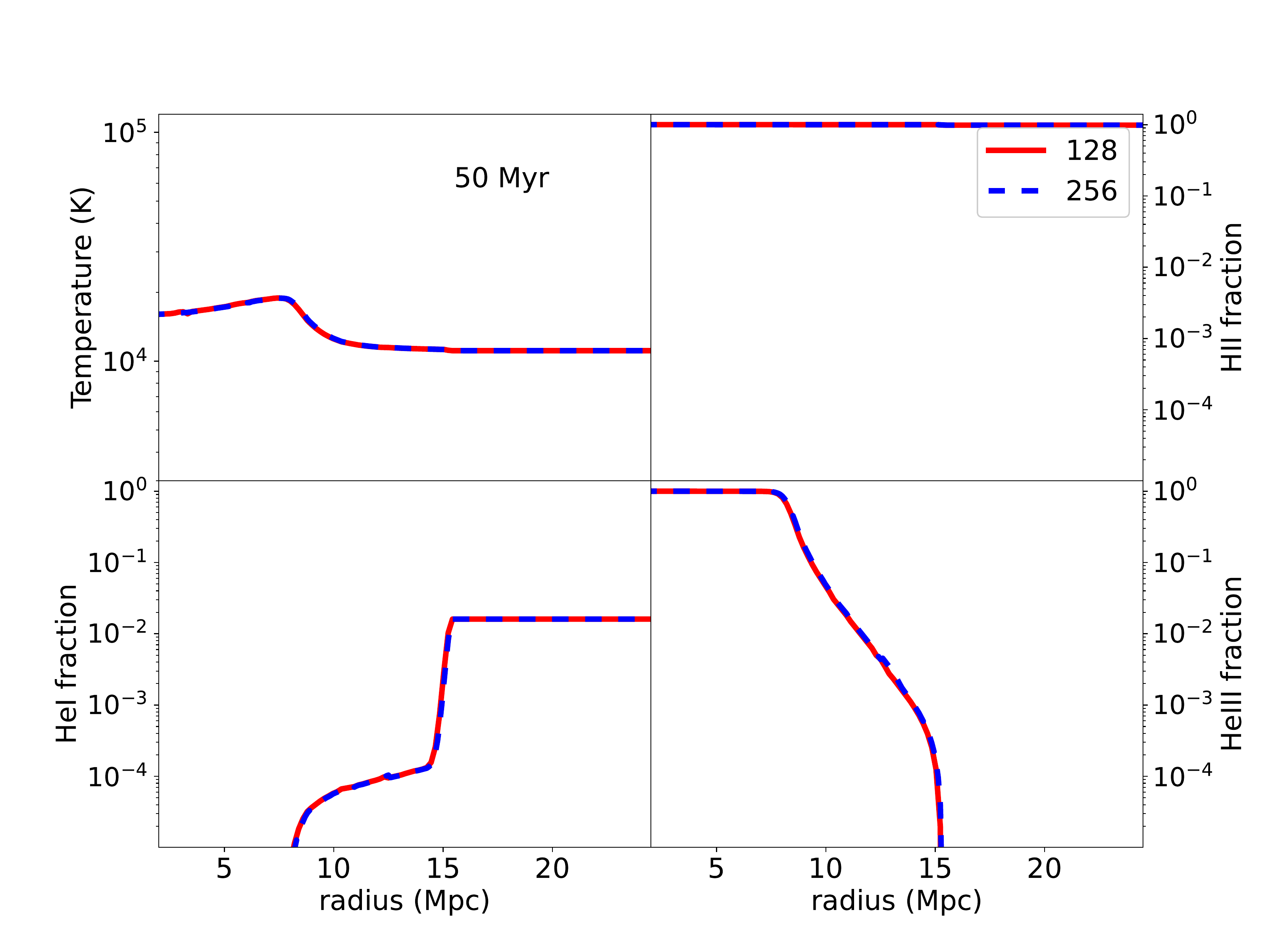}
    \caption{Profiles at 50 Myr for reionization at $z=4$ by a source $L_{S,\nu}\sim\nu^{-1.73}$. All the quantities are converged in these $(25$ Mpc, $128^{3})$ and $(25$ Mpc, $256^{3})$ simulations.}
    \label{fig:QSO_1.7_4}
\end{figure}

The convergence test results are shown in Figs.~\ref{fig:QSO_0.5_6} - \ref{fig:QSO_1.7_4}, for the power-law test problems for IGM mean densities at $z=6$ and 4, and for spectral indices $\alpha_Q=0.5$ and 1.73. Convergence is generally reached in the inner ionised regions for the $128^3$ simulations, but, particularly for the softer spectrum, convergence in temperature and the ionization structure is improved on going to $256^3$. The latter corresponds to an initial optical depth per cell at the hydrogen photoelectric threshold of 124 at $z=6$ and an initial optical depth per cell at the singly ionised helium photoelectric threshold of 0.9 at $z=4$.


\section{Revisions to \texttt{ENZO}}
\label{appendix:enzo_revision}

We describe revisions to \texttt{ENZO} v2.6 to implement photoionization by a central source.

We check the consistency of the source codes, especially the consistency of the codes relevant to the ray-tracing module. We find bugs in the implementation in \texttt{ENZO} v2.6 significantly affect the accuracy of the results. These are demonstrated in Appendix~\ref{appendix:Ionizedball} by simulating a classical ray-tracing problem, the formation of a Str\"omgren sphere \citep{2006MNRAS.371.1057I}.

We impose new methods and restrictions on the ray-tracing module to make \texttt{ENZO} suitable for both static and cosmological hydrodynamical simulations with high-luminosity radiation sources. The modifications are:\ a)\  Probabilistic Absorption Method (Eqs.~[\ref{eq:abs_HI}]--[\ref{eq:abs_HeII}]); b)\ \HeIII\ Ionisation Adaptive Time Step Scheme (Appendix~ \ref{appendix:HeIII_ionisation}); c)\ Restriction on Photon Package Travel Distance (Appendix~\ref{appendix:horizon}). 

\subsection{Test problem:\ Str\"omgren sphere}
\label{appendix:Ionizedball}
A Str\"omgren sphere
is the final stage of an isotropically expanding ionization region with a central source in a uniform medium once ionizations are balanced by recombinations. As a test problem, the Str\"omgren sphere simulation has a few key benefits:\ a) the solution is analytical, hence it is easy to check the accuracy of the results; b) the solution is isotropic; as a result, any artificial inhomogeneity caused by the algorithm is visible. The analytic solution for the radius of the ionisation front is:
\begin{equation}
 R(t) = R_S \left[1 - \exp\left(-\frac{t}{t_{\rm rec}}\right) \right]^{\frac{1}{3}},
	\label{eq:sphere_H}
\end{equation}
where $R_S = (3\dot N_\gamma/4\pi n_\mathrm{H} n_e \alpha)^{1/3}$ is the final radius of the ionised region (the Str\"omgren radius), $\dot N_\gamma$ is the photon emission rate, $n_{\rm H}$ is the hydrogen number density, $n_e$ is the electron number density, $\alpha$ is the radiative recombination rate within the ionised region and $t_{\rm rec} = 1/n_e \alpha$ is the recombination time (assumed constant).

We adopt a similar parameter set to that used in \cite{2006MNRAS.371.1057I} to compare with the results therein. Specifically, the monochromatic source emits photons at the rate $\dot N_\gamma = 5 \times 10^{48}\,\mathrm{s^{-1}}$, the simulation box size is $6.6\,\mathrm{kpc}$ and the total number of cells is $128^3$, the minimum ray angular resolution parameter is $\Phi_c = 5.1$ \citep[see the definition in][] {2011MNRAS.414.3458W}, the number density of hydrogen atoms is $n_\mathrm{H} = 10^{-3}\,\mathrm{cm^{-3}}$ and the recombination rate is $\alpha = 2.59 \times 10^{-13}\,\mathrm{ cm^3 \ s^{-1}}$ at $T = 10^4$~K, leading to $R_S = 5.4\,\mathrm{kpc}$ and $t_{\rm rec} = 122\,\mathrm{Myr}$. We note that in \cite{2006MNRAS.371.1057I}, the temperature is fixed at $T = 10^4$~K; however, \texttt{ENZO} does not support this setting due to its formulation of the internal chemistry and energy solvers. To make the simulation close to the analytical problem, we adjust the energy of the monochromatic photon to $23.26\,\mathrm{eV}$ and set the adiabatic index to $\gamma = 1.667$ to ensure that the gas temperature in the ionized region is close to $T = 10^4$~K. This approximation and these parameters keep the maximum temperature deviation to within $20\%$ in all Str\"omgren sphere simulations. This ensures the Str\"omgren radius will be matched to better than $4\%$.

\begin{figure}
    \includegraphics[width=\columnwidth]{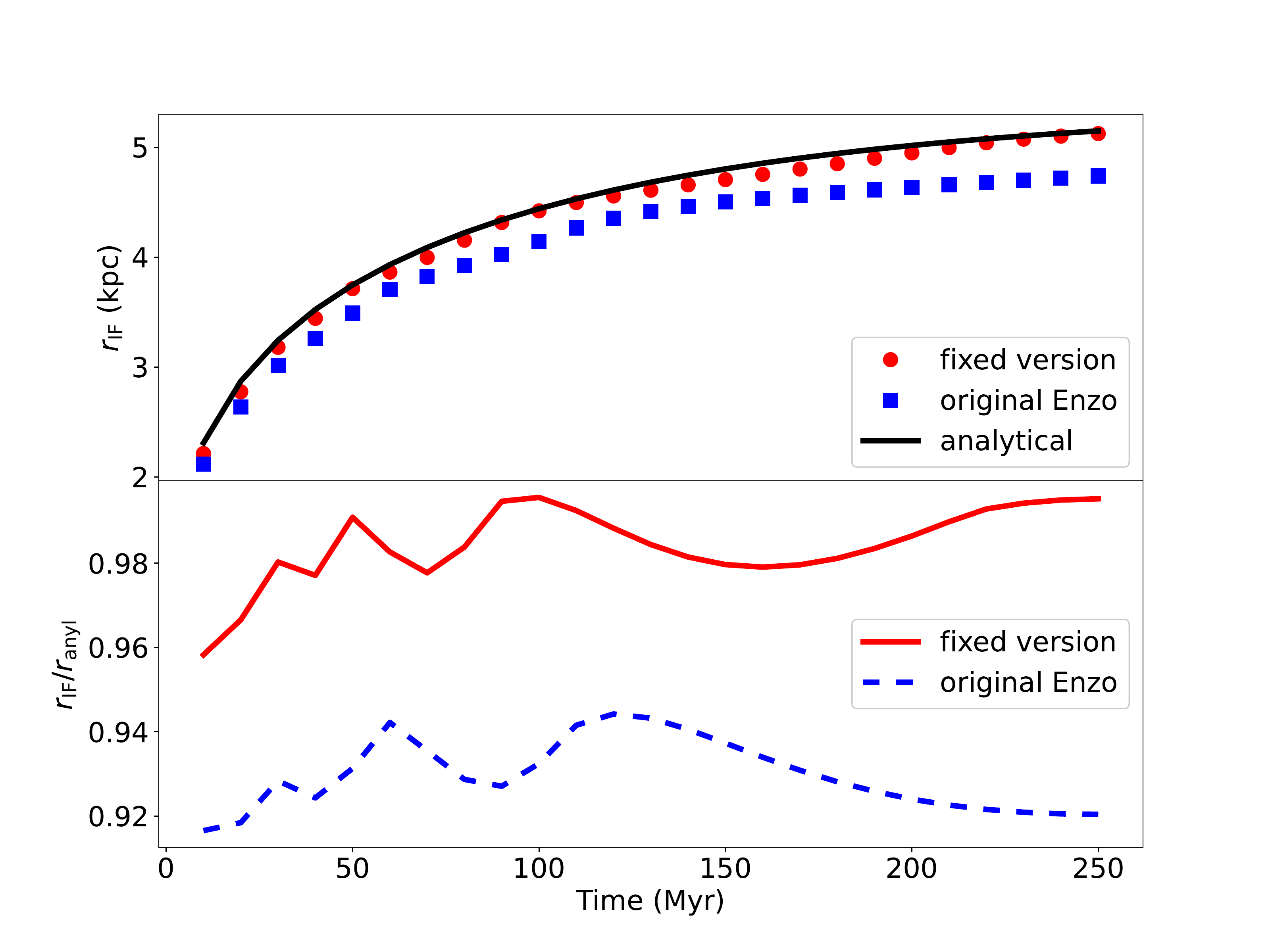}
    \caption{In the upper panel, the black solid line shows the analytic solution for the time development of the ionization radius. The red circles show the ionization radii computed by \texttt{ENZO} with our revisions, and the blue squares are the results from \texttt{ENZO} v2.6. The lower panel shows the ratio of the computed and analytic solutions for the ionization radius. The modifications significantly improve agreement with the analytic solution.
}
    \label{fig:HI_front}
\end{figure}

Figure \ref{fig:HI_front} shows the evolution of the ionization radius computed by our revised version of \texttt{ENZO} and by \texttt{ENZO} v2.6. The result from our version very closely matches the analytical result, validating our revisions to \texttt{ENZO} v2.6.

\subsection{\HeIII\ Ionisation Adaptive Time Step Scheme}
\label{appendix:HeIII_ionisation}
We include the maximum changing rate of the \HeIII\ fraction as an additional condition on the ray-tracing adaptive time step scheme in \texttt{ENZO}. The original ray-tracing adaptive time step in \texttt{ENZO} is based on the maximum changing rate of the \HII\ fraction \citep[see][for more details]{2011MNRAS.414.3458W}. However, during the helium reionization epoch, the hydrogen atoms in the IGM have already been fully ionised.
As a consequence, hydrogen is optically thin to the ionising photons and the \HII\ fraction changes very slowly. By contrast, the ionisation fraction of \HeII\ is rapidly changing in this period, as the gas is initially optically thick to \HeII\ ionising photons. Therefore, an additional restriction based on the maximum changing rate of \HeIII\ is implemented in our simulations.

We also extend the applicability of the ray-tracing adaptive time step scheme:\ the original algorithm of the time step calculator does not consider the influence caused by the expansion of the Universe. This feature leads to overestimating the ray-tracing time steps in cosmological simulations, bringing artificial effects into cosmological simulations.  


\subsection{Restriction on Photon Package Travel Distance}
\label{appendix:horizon}
The comoving light travel distance from the radiation sources is used to avoid photon packages from moving beyond the particle horizon of the photons in our implementation. Similarly to other ray-tracing algorithms in cosmological simulation codes \citep{2002MNRAS.330L..53A, 2006MNRAS.371.1057I}, the ray-tracing algorithm in \texttt{ENZO} assumes the propagation speed of the photon package is infinite at every radiation transfer (RT) time step. This approximation is used for the following reasons: 

a)\ It is computationally cheaper. In a simulation with a radiation source, where the propagation speed is finite, the total number of photon packages, which are stored in system memory (such as RAM) between two RT time steps, is proportional to the volume of the ionized region. Also, in such an algorithm, various information about a photon package, such as its position, direction, luminosity and birth time, are required to trace the photon package. Hence, assuming a finite propagation speed significantly increases the cost to system RAMs and makes the simulation computationally prohibitive. On the other hand, in a simulation with an infinite speed of light, all the photon packages are generated and deleted at every RT time step. Thus, the photon packages do not occupy any system RAM in between two RT time steps. As a consequence, the infinite propagation speed approximation is more practical from a technical point of view. This instantaneous RT approximation amounts to neglecting the time differential operator on the left hand side of Eq.~(\ref{eq:tdRT}).

b)\ The expansion speed of the ionisation bubbles is generally much slower than the speed of light. Also, most of the ionising photons associated with a ray are absorbed in the ionisation front. Therefore, when the particle horizon is distant from the ionisation front, the influence caused by the infinite propagation speed approximation is marginal. 

Although assuming the infinite propagation speed is a safe approximation in most simulated situations, the particle horizon needs to be imposed as the maximum travel distance of rays in many situations. These include:\ the radius of the ionised bubble could be comparable to the particle horizon when the source just begins to shine; in a large-scale simulation, the box size of the simulation could be larger than the particle horizon during the simulation period. In these situations, forbidding photon packages from transferring across their corresponding particle horizons also reduces the computation time. In our revised version of \texttt{ENZO}, we delete a photon packet if it reaches its particle horizon.    

\label{lastpage}

\end{document}